\documentclass[useAMS,usenatbib]{mn2e}

 \usepackage{times}
 \usepackage{graphicx}

\def\pcm3{{\rm\thinspace cm$^{-3}$}}

\def\contcaption{\@conttrue\SFB@caption\@captype}

\title[Astrometric and photometric IMF of USco]
{Astrometric and photometric initial mass functions from the UKIDSS Galactic Clusters Survey: IV Upper Sco
\thanks{Based on observations made with the United Kingdom Infrared
Telescope, operated by the Joint Astronomy Centre on behalf of the
U.K. Particle Physics and Astronomy Research Council.}}
\author[N. Lodieu et al.]{N. Lodieu$^{1,2}$\thanks{E-mail: nlodieu@iac.es} \\
$^{1}$Instituto de Astrof\'isica de Canarias (IAC), V\'ia L\'actea s/n,
E-38205 La Laguna, Tenerife, Spain \\
$^{2}$ Departamento de Astrof\'isica, Universidad de La Laguna (ULL),
E-38205 La Laguna, Tenerife, Spain} 
\begin{document}

\date{Accepted \today{}. Received \today{}; in original form \today{}}

\pagerange{\pageref{firstpage}--\pageref{lastpage}} \pubyear{2005}

\maketitle

\label{firstpage}

\begin{abstract} 
We present the results of a proper motion wide-field near-infrared 
survey of the entire Upper Sco (USco) association ($\sim$160 square 
degrees) released as part of the UKIRT Infrared Deep Sky (UKIDSS) Galactic 
Clusters Survey (GCS) Data Release 10 (DR10). We have identified a 
sample of $\sim$400 astrometric and photometric member candidates 
combining proper motions and photometry in five near-infrared passbands
and another 286 with $HK$ photometry and 2MASS/GCS proper motions. 
We also provide revised membership for all previously published USco 
low-mass stars and substellar members based on our selection and
identify new candidates, including in regions affected by extinction.
We find negligible variability between the two $K$-band epochs, below
the 0.06 mag rms level. We estimate an upper limit of 2.2\% for wide 
common proper motions with projected physical separations less than 
$\sim$15000 au. We derive
a disk frequency for USco low-mass stars and brown dwarfs between
26 and 37\%, in agreement with estimates in IC\,348 and $\sigma$\,Ori.
We derive the mass function of the association and find it consistent
with the (system) mass function of the solar neighbourhood and other
clusters surveyed by the GCS in the 0.2--0.03 M$_{\odot}$ mass range.
We confirm the possible excess of brown dwarfs in USco.
\end{abstract}


%
%
\section{Introduction}
The knowledge of the number of stars and brown dwarfs as a function
of mass in open clusters and star-forming regions is important to
address the question of the universality of the initial mass function
\citep{salpeter55,miller79,scalo86,kroupa02,chabrier03,kroupa11}.
The advent of large optical and near-infrared detectors has shed
light on the properties of low-mass stars and substellar objects in
a variety of environments and enabled an in-depth study of the
mass function well below the hydrogen-burning limit 
\citep[see review by][and references therein]{bastian10}.
However, many surveys in young regions lack homogeneity in the 
multi-band photometric coverage and accurate proper motions for brown 
dwarf members, making interpretation of their mass spectrum sometimes
difficult.

The UKIRT Infrared Deep Sky Survey 
\citep[UKIDSS;][]{lawrence07}\footnote{The survey is described at 
www.ukidss.org} is a deep large-scale infrared survey conducted with 
the UKIRT Wide field CAMera \citep[WFCAM;][]{casali07} equipped 
with five infrared filters \citep[$ZYJHK$;][]{hewett06}. All data are 
pipeline-processed at the Cambridge Astronomical Survey Unit 
\citet[CASU;][Irwin et al.\ in preparation]{irwin04}\footnote{The 
CASU WFCAM webpage is at
http://apm15.ast.cam.ac.uk/wfcam}, processed and archived in Edinburgh, 
and later released to the community through the WFCAM Science Archive 
\citep[WSA;][]{hambly08}\footnote{WSA is accessible
at http://surveys.roe.ac.uk/wsa}.
One of its components, the Galactic Clusters Survey (hereafter GCS) 
imaged $\sim$1000 square degrees homogeneously in ten star-forming 
regions and open clusters down to 0.03--0.01 M$_{\odot}$ (depending on 
the age and distance of each region) to investigate the universality of 
the initial mass function. In addition to the photometry, the latest 
releases of the GCS provide proper motions measured from the different 
epochs, with accuracies of about five per year (mas/yr).

The USco region is part of the nearest OB association to the Sun, Scorpius
Centaurus, located at 145 pc \citep{deBruijne97}. Its precise age is
currently under debate \citep{song12}: earlier studies using isochrone
fitting and dynamical studies derived an age of 5$\pm$2 Myr 
\citep{preibisch02} in agreement with deep surveys 
\citep{slesnick06,lodieu08a} but recently challenged by \citet{pecaut12} 
who quoted 11$\pm$2 Myr from a spectroscopic study of F stars at optical
wavelengths. The association was targeted at multiple 
wavelengths, starting off in X~rays \citep{walter94,kunkel99,preibisch98}, 
but also astrometrically with Hipparcos \citep{deBruijne97,deZeeuw99}, 
and more recently in the optical 
\citep{preibisch01,preibisch02,ardila00,martin04,slesnick06} and in the 
near--infrared \citep{lodieu06,lodieu07a,dawson11,lodieu11a,dawson12}.
Tens of brown dwarfs have now been confirmed spectroscopically as 
USco members 
\citep{martin04,slesnick06,lodieu06,slesnick08,lodieu08a,martin10a,dawson11,lodieu11a}
and the mass function of this population determined well into the 
substellar regime \citep{slesnick08,lodieu11a}. Three independent 
studies noticed that USco may harbour an excess of brown dwarfs 
\citep{preibisch01,lodieu07a,slesnick08}. Five T--type candidates
reported by \citet{lodieu11c} have been rejected as astrometric 
members of the association (Lodieu et al.\ 2013, in press).

In this paper we present a photometric and proper motion-based study of 
$\sim$50 square degrees in USco released as part of the UKIDSS GCS DR10 
(14 January 2013) along with a revised analysis of the GCS Science
Verification (SV) data (6.7 square degrees). This study is complemented 
by $HK$ imaging and proper motion from the 2MASS/GCS cross-match 
for the remaining area of the association.
Our work improves on previous studies by selecting members based on
accurate proper motions provided by the GCS down to masses as low as
0.01 M$_{\odot}$ and identifying candidates in regions previously
unstudied and affected by heavy extinction.
In Section \ref{USco:sample} we present the photometric and astrometric
dataset employed to extract USco member candidates.
In Section \ref{USco:status_old_cand} we review the list of previously 
published USco members recovered by our analysis and revise 
their membership.
In Section \ref{USco:new_cand}  we identify new stellar and substellar member
candidates based on five-band photometry and astrometry.
In Section \ref{USco:variability} we investigate the level of $K$-band variability 
for USco low-mass stars and brown dwarfs.
In Section \ref{USco:IMF} we derive the cluster luminosity and 
(system) mass functions and compare it to earlier estimates for this 
cluster and others, along with that of the field population.
This work is in line with our recent studies of the Pleiades \citep{lodieu12a},
$\alpha$\,Per \citep{lodieu12c}, and Praesepe \citep{boudreault12} clusters.

%
%
\begin{figure}
   \centering
   \includegraphics[width=\linewidth]{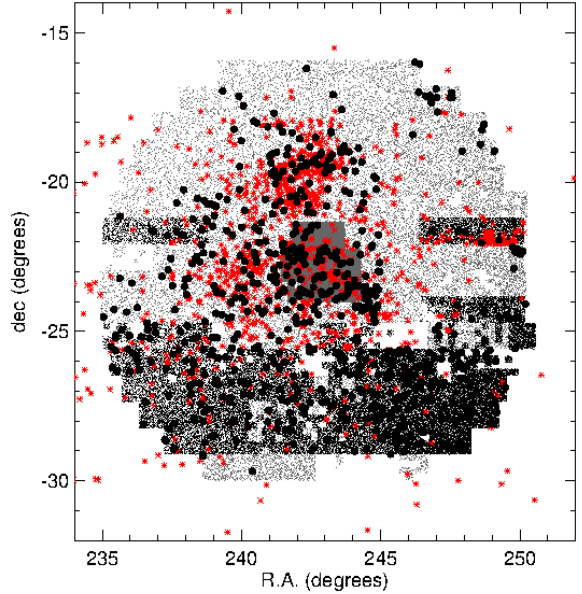}
   \caption{The coverage in USco as released by the UKIDSS GCS DR10:
the light grey, dark grey, and black patches indicate the $HK$, SV,
and GCS DR10 samples, respectively. 
The holes are due to frames removed from the GCS 
release due to quality control issues. Overplotted are member candidates 
identified in this work (filled black dots) and previously-published
sources from the literature (red asterisks).
}
   \label{fig_USco:coverage_DR10}
\end{figure}
%

%
%
\begin{figure}
   \includegraphics[width=1.00\linewidth]{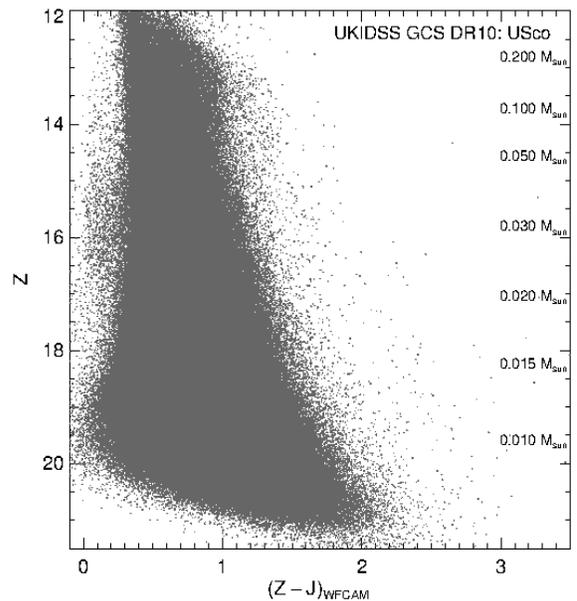}
   \caption{($Z-J$,$Z$) CMD for $\sim$50 square degrees in USco 
extracted from the UKIDSS GCS DR10\@. The mass scale shown 
on the right hand side spans $\sim$0.2--0.08 M$_{\odot}$, following 
the 5-Myr BT-Settl isochrones \citep{allard12}.
}
   \label{fig_USco:ZJZcmd_alone_DR10}
\end{figure}
%

%
%
\begin{table}
  \centering
  \caption{Approximate coordinates of the USco regions from the 
$ZYJHK-$PM sample with and without extinction, the GCS SV, and 
the $HK$-only coverage (one in 100 source shown; 
Fig.\ \ref{fig_USco:coverage_DR10}).
}
  \label{tab_USco:regions}
  \begin{tabular}{l c c c}
  \hline
Region            &  R.A. & dec   & Area  \cr
  \hline
                  &  deg  & deg   & deg$^{2}$ \cr
  \hline
No Extinction \#1    & $\leq$244.4   &  any                & 33.3   \cr
No Extinction \#2    & $\geq$244.4   &  $>$\,$-$22.5       &  3.3   \cr
Extinction \#1       & 244.4--248.0  &  $\leq$\,$-$27      &  7.7   \cr
Extinction \#2       & 244.0--246.0  &  $-$27.0 to $-$24.5 &  3.0   \cr
Extinction \#3       & $\geq$246.0   &  $-$26.0 to $-$23.5 &  5.7   \cr
SV \#1               & 241.4--244.3  &  $-$24.0 to $-$22.2 &  5.2   \cr
SV \#2               & 241.8--243.7  &  $-$22.2 to $-$21.4 &  1.5   \cr
$HK$ with Extinction & 245.0--249.5  &  $-$25.6 to $-$19.0 & 29.7   \cr
 \hline
\end{tabular}
\end{table}

%
%
\begin{figure*}
   \includegraphics[width=0.495\linewidth]{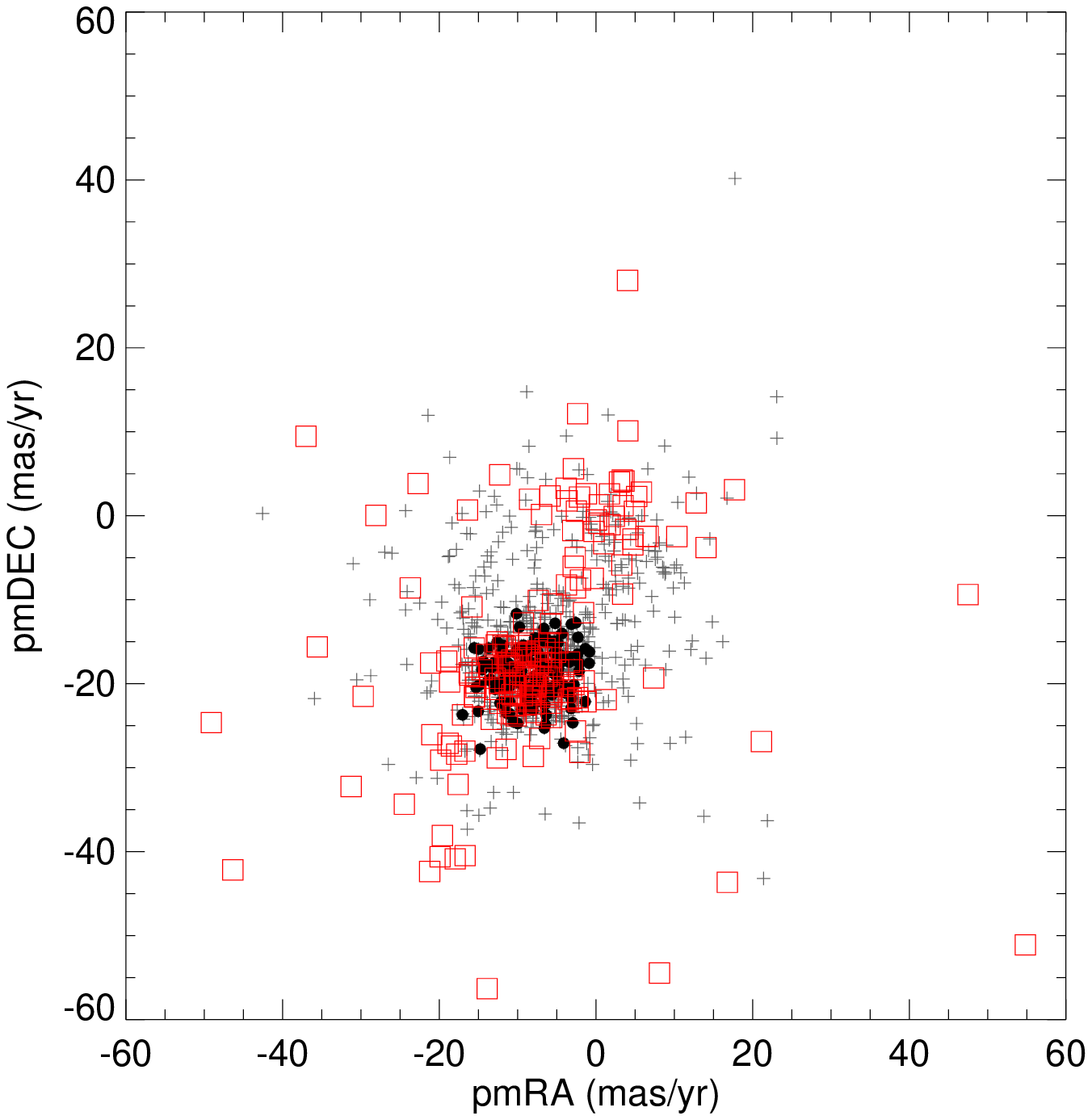}
   \includegraphics[width=0.495\linewidth]{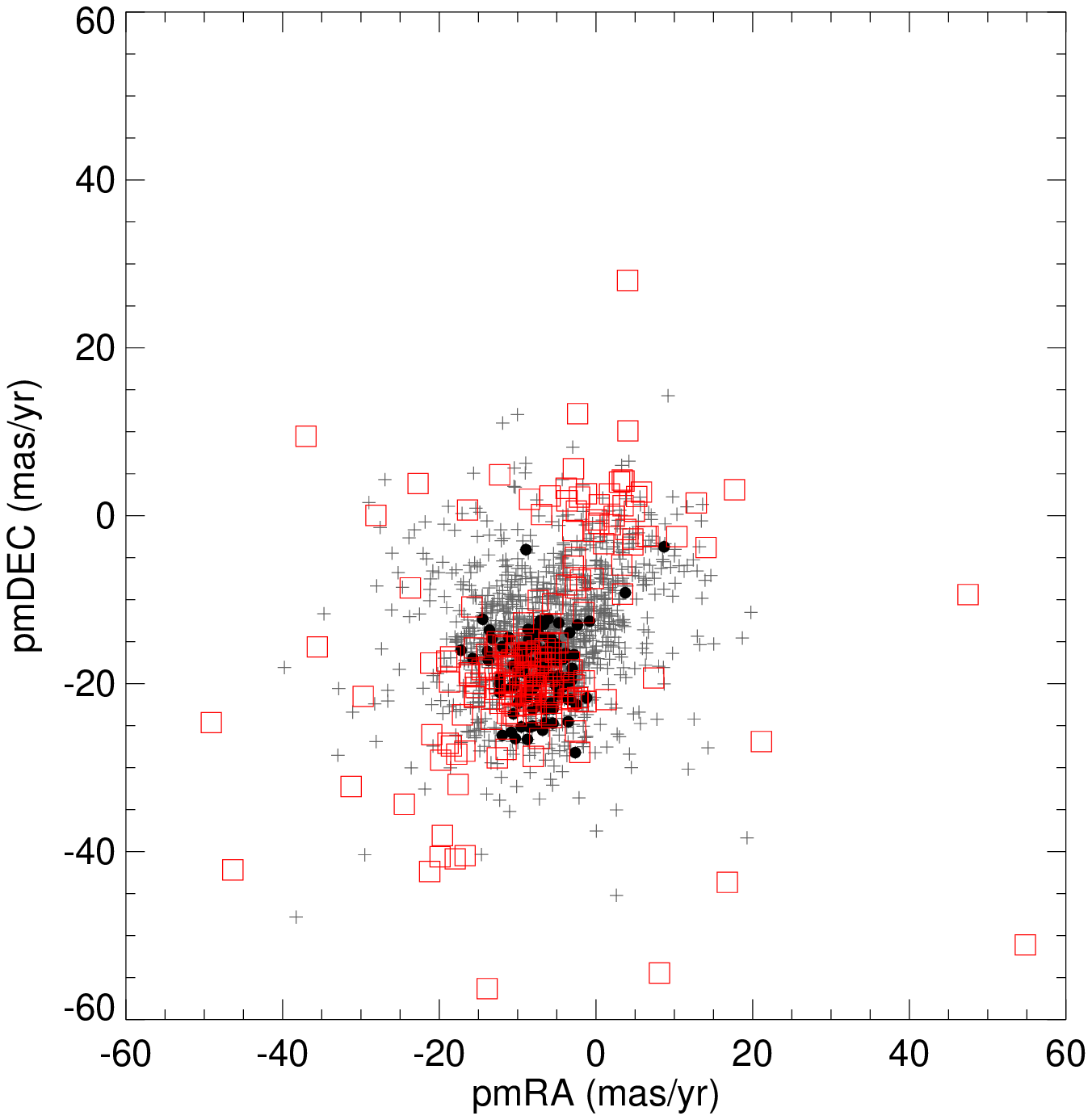}
   \caption{Vector point diagrams showing the proper motions from the
   GCS alone in right 
ascension (x-axis) and declination (y-axis) for previously-known member 
candidates recovered by the GCS DR10 (red open squares) and all point
sources after the crude photometric selection made in the ($Z-J$,$Z$)
colour-magnitude diagram for the $ZYJHK-$PM sample. Black filled dots 
are our photometric and astrometric member candidates in USco.
{\it{Left:}} Vector point diagram for a region without extinction.
{\it{Right:}} Same diagram for the area of USco affected by reddening.
}
   \label{fig_USco:diagram_VPD}
\end{figure*}
%

%
%
\section{The sample}
\label{USco:sample}

We selected point sources in the full USco region, defined by 
RA=230--252 degrees and declinations between $-$32 and $-$16 degrees 
(Fig.\ \ref{fig_USco:coverage_DR10}). We retrieved the catalogue using 
a Structure Query Language (SQL) query similar to our earlier studies 
of the Pleiades \citep{lodieu12a}, $\alpha$\,Per \citep{lodieu12c},
and Praesepe \citep{boudreault12}. Briefly, we selected high quality
point sources with $JHK$ photometry, allowing for $Z$ and $Y$ non
detections. The query returned a total number of 2,943,321 sources.
We refer to this sample as the ``$ZYJHK-$PM''sample throughout the
paper. Below we distinguish the region free of extinction and the one 
affected by reddening although we will show that the same photometric 
and astrometric criteria can be applied to provide a clean sample of
member candidates.

Proper motion measurements are available in the WFCAM Science Archive 
for UKIDSS data releases from DR9 for all the wide/shallow surveys with 
multiple epoch coverage in each field (i.e.\ the LAS, GCS and GPS). 
Details of the procedure are in \citet{collins12} and summarised 
in \citet{lodieu12a} for the purpose of the Pleiades. The typical error
bars on the GCS proper motions in USco are 4 mas/yr and 6 mas/yr, down to 
$Z$\,=\,19 mag and 20 mag, respectively (Fig.\ \ref{fig_USco:diagram_VPD}).

\begin{figure*}
   \centering
   \includegraphics[width=0.495\linewidth]{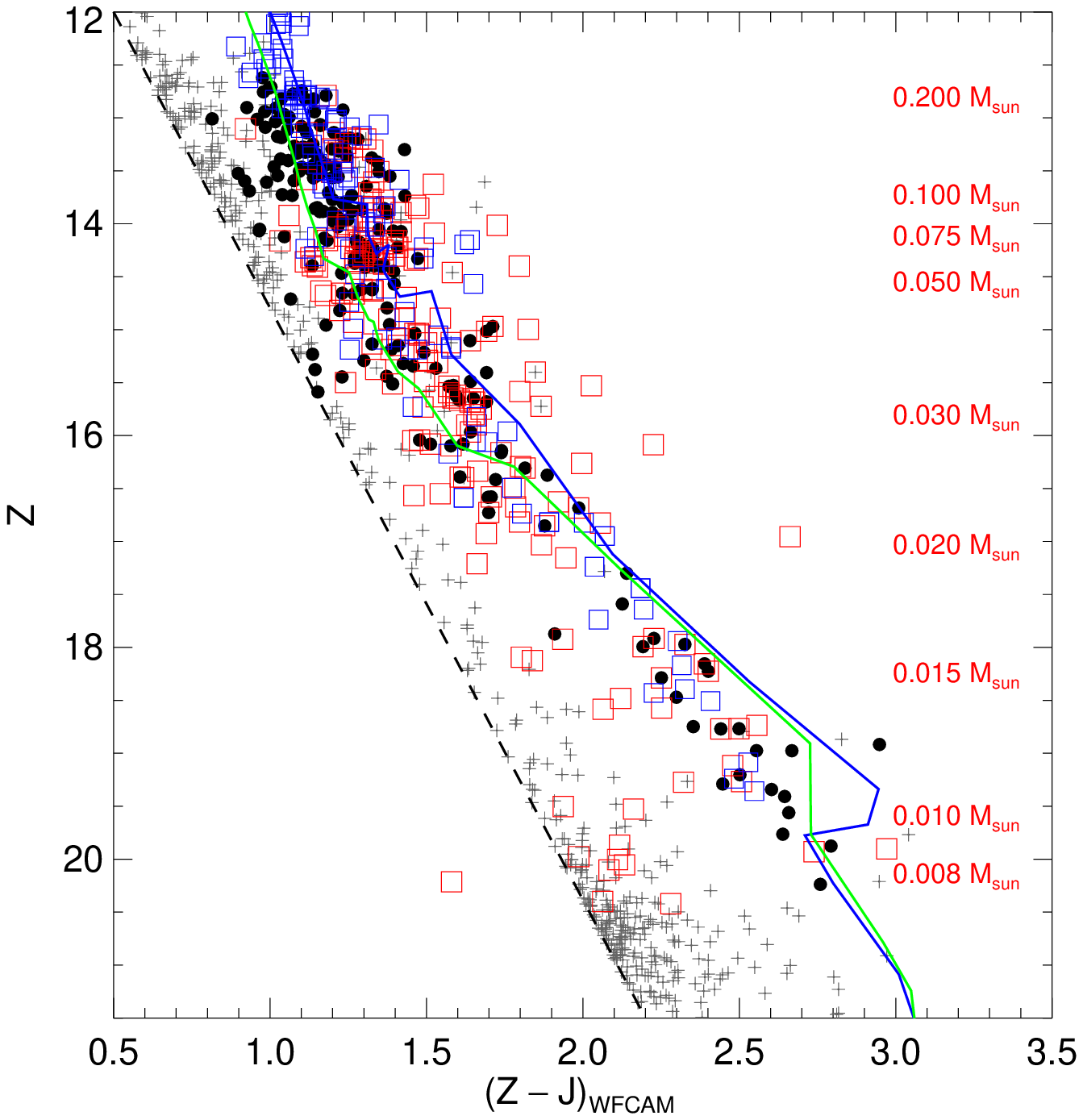}
   \includegraphics[width=0.495\linewidth]{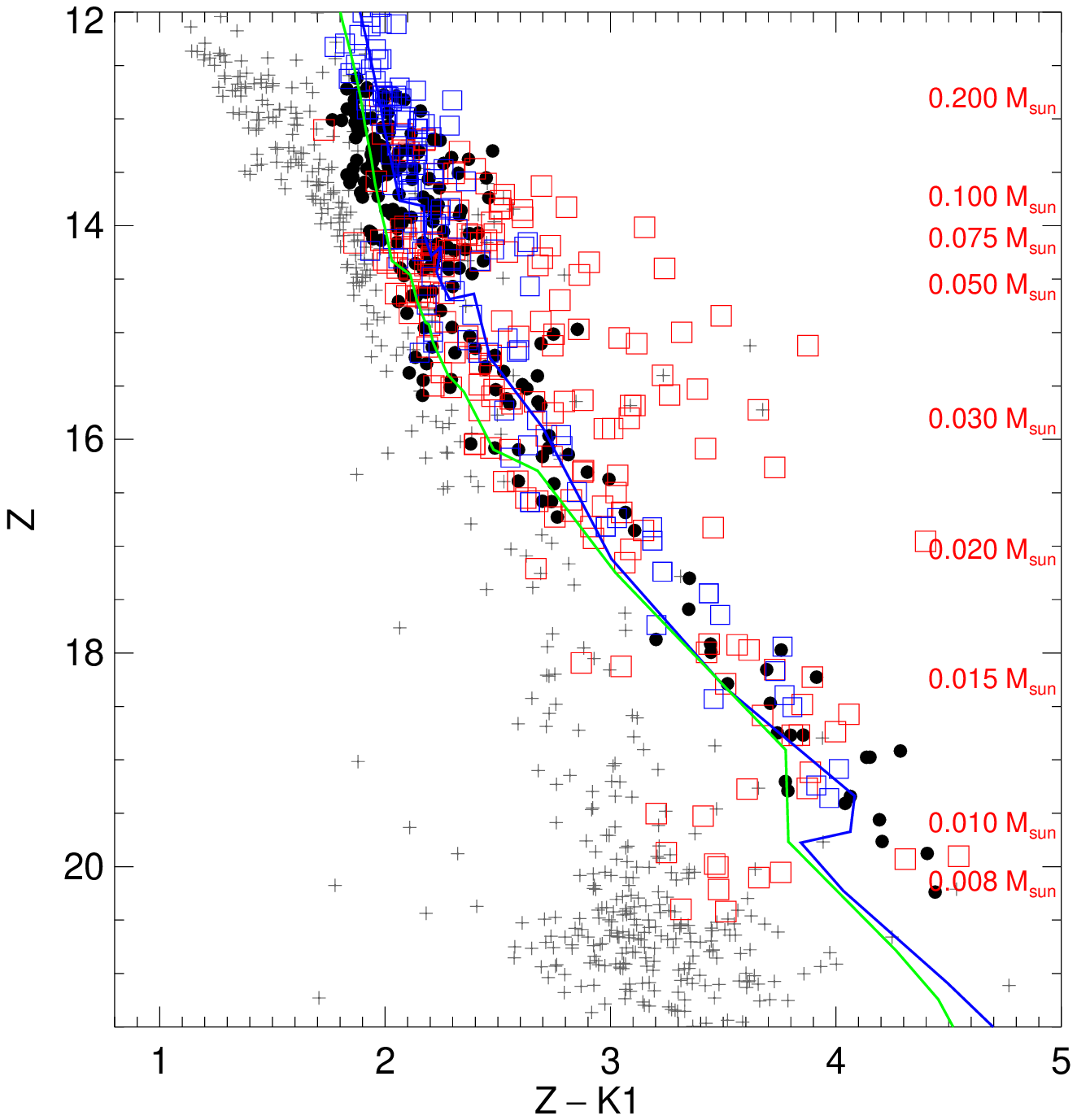}
   \includegraphics[width=0.495\linewidth]{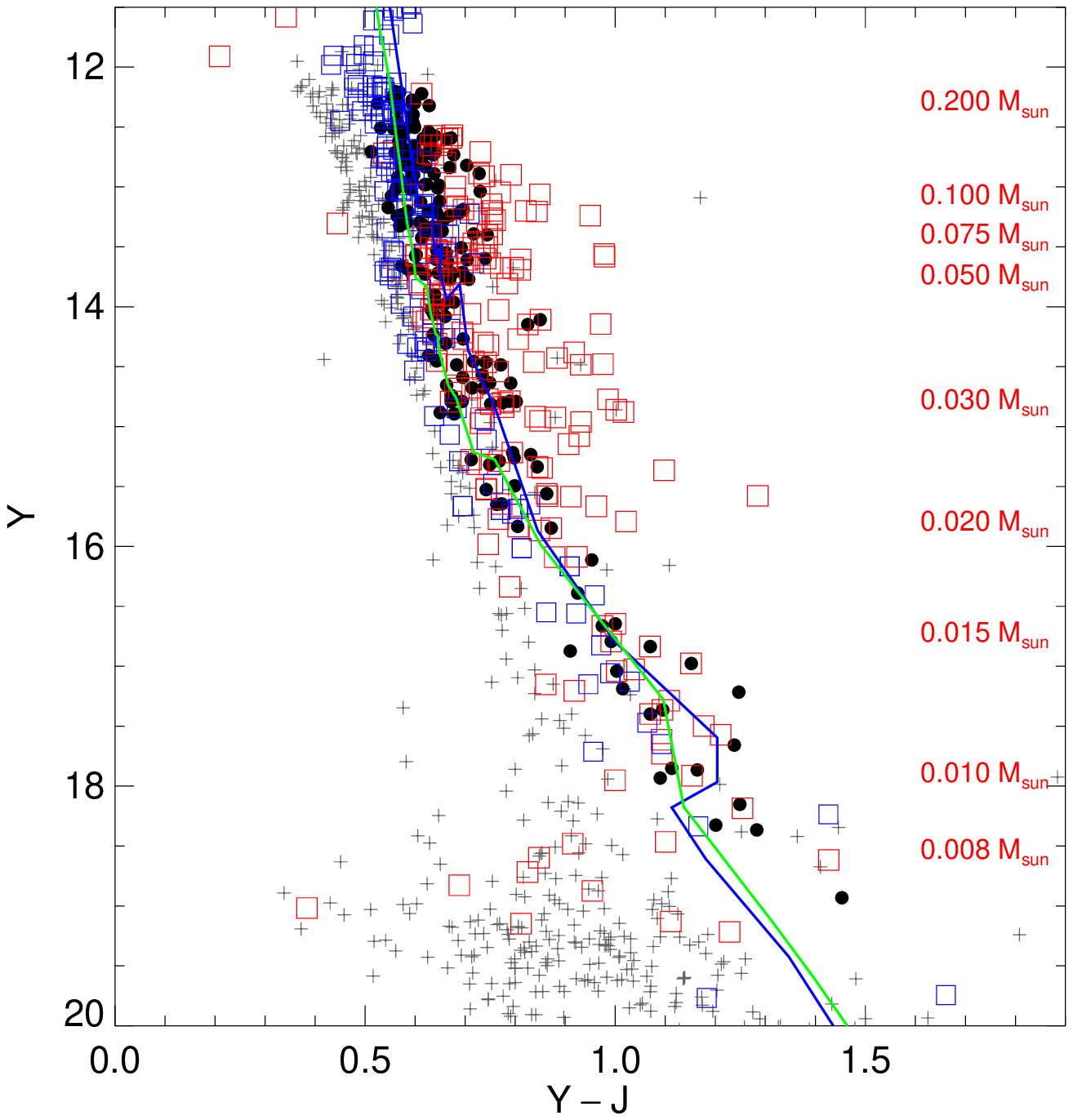}
   \includegraphics[width=0.495\linewidth]{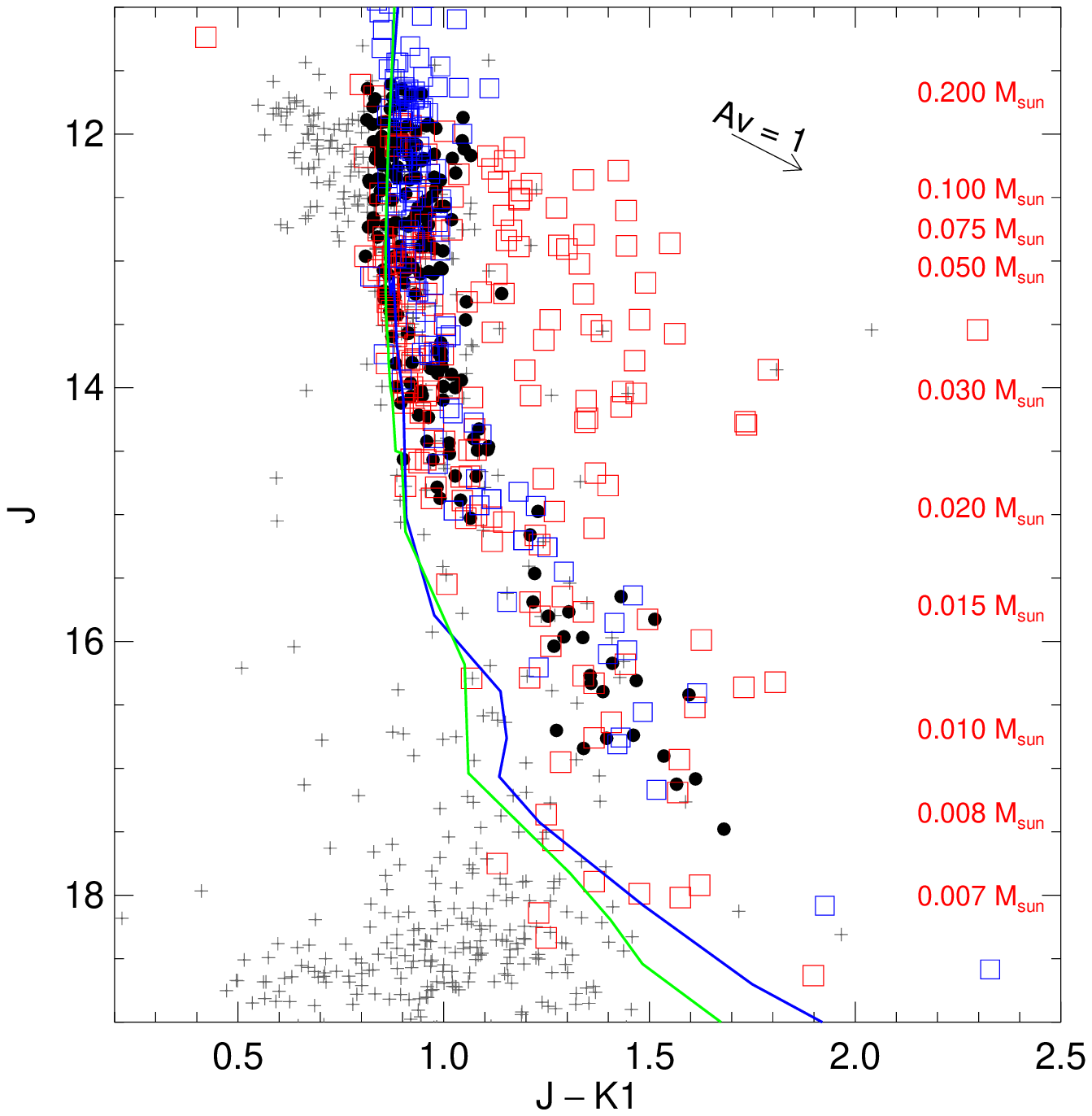}
   \caption{Colour-magnitude diagrams showing the USco member candidates
previously reported in the literature (red open squares) and the new ones 
identified in this work (black dots). Photometric and/or proper motion
non-members are highlighted as grey crosses. Known spectroscopic members
are overplotted as blue open squares \citep{lodieu11a}.
Overplotted are the 5 and 10 Myr-old BT-Settl isochrones \citep{allard12}
shifted at a distance of 145 pc. The mass scale shown on the right hand 
side of the diagrams spans approximately 0.2--0.008 M$_{\odot}$, 
according to the 5 Myr isochrones.
The dashed {\bf{line}} in the upper left diagram represent our crude photometric
selection using a line running from ($Z-J$,$Z$)\,=\,(0.5,12.0) to (2.2,21.5).
   {\it{Upper left:}} ($Z-J$,$Z$);
   {\it{Upper right:}} ($Z-K$,$Z$);
   {\it{Lower left:}} ($Y-J$,$Y$);
   {\it{Lower right:}} ($J-K$,$J$).
}
   \label{fig_USco:YJK_cmds}
\end{figure*}

First, we applied a crude photometric selection to work with a subsample
of the entire catalogue. We selected all sources located to the right
of the line running from (0.5,12) to (2.2,21.5) in the ($Z-J$,$Z$) 
colour-magnitude diagram (Fig.\ \ref{fig_USco:YJK_cmds}). 
We made sure that this line allowed us to recover known spectroscopic 
members (see Section \ref{USco:status_old_cand}). This sample contains 
29,382 sources, divided into 9351 in the region free of extinction 
and 20,031 in the parts affected by reddening 
(Fig.\ \ref{fig_USco:coverage_DR10}; Table \ref{tab_USco:regions}).

The rest of the USco association is not covered with enough epochs to
measure proper motions based only on GCS data. This is the case for the 
GCS SV (Table \ref{tab_USco:regions}) and the area covered in $HK$ only.
In the case of the GCS SV area, we have $ZYJHK$ photometry and proper
motions measured from the 2MASS/GCS cross-match. \citet{lodieu07a}
identified member candidates in this part of the association (although
with a slightly smaller area released at that time) and 
confirmed a large number as spectroscopic members \citep{lodieu11a}.
\citet{dawson12} also included this region in their study of the disk
properties of USco low-mass and brown dwarfs, as did \citet{riaz12a}.
We have 430 sources in the GCS SV region, after applying the crude
photometric selection described above.

We added to those samples the full coverage of the GCS DR10, only
imaged in the $H$ and $K$ passbands (hereafter the $HK$ sample).
We also measured proper motions from the 2MASS/GCS correlation.
Our query returned a total of 7,328,848 to which we should remove
the GCS SV and GCS DR10 samples as well as the region
most affected by reddening (defined by R.A.\,=\,245--249.5 degrees and 
dec between $-$25.6 and $-$19 degrees, see Table \ref{tab_USco:regions}).
We applied a crude photometric selection in the ($H-K$,$H$) 
colour-magnitude diagram, keeping only sources to the right of a line
running from ($H-K$,$H$)\,=\,(0.2,12) to (0.7,18). We are left with
39,450 sources to investigate astrometrically 
(Sect.\ \ref{USco:new_cand_PM}).

%
%
\section{Cross-match with previous surveys}
\label{USco:status_old_cand}

We compiled a list of USco members published over the past
decades by various groups 
\citep{preibisch98,preibisch01,preibisch02,ardila00,martin04,slesnick06,lodieu06,lodieu07a,lodieu08a,dawson11,lodieu11a,dawson12,luhman12c}
to update their membership status with the photometry and astrometry 
provided by the GCS DR10 (Table \ref{tab_USco:early_summary}). 
This list will serve as starting point to identify new member candidates
in the GCS, estimate the mean (relative) proper motion of USco members,
and derive the cluster luminosity and mass functions. 
We compiled a list of 2079 candidates, reduced to 1566 after removing
multiple pairs.

We cross-correlated this list of 1566 known member candidates with the
$ZYJHK-$PM catalogue using a matching radius of three arcsec and 
found 125 sources in common (red open squares in 
Fig.\ \ref{fig_USco:diagram_VPD}; Table \ref{tab_USco:early_summary}). 
We repeated the same process with the GCS SV and $HK$-only areas,
yielding 73 and 651 member candidates in common, respectively
(Table \ref{tab_USco:early_summary}).
The number of known member candidates recovered in GCS DR10 is 
generally low because most surveys focussed on the northern area with 
right ascensions between 240 and 245 degrees and declinations above 
$-$25$^{\circ}$ \citep[see list of sources in][]{luhman12c}. Below we 
provide a few comments on the recovery rate for the early studies 
listed in Table \ref{tab_USco:early_summary}.
\begin{itemize}
\item The samples published by \citet{preibisch01} and \citet{preibisch02}
lie outside the GCS DR10 and SV areas and very few objects of the X-ray
and proper motion catalogues of \citet{preibisch98} lie in those regions
as well {\bf{as}} X-ray and proper motion samples of \citet{preibisch98}.
Most of the members from \citet{preibisch98} and \citet{preibisch01} are 
too bright for the UKIDSS GCS and generally saturated because they are 
brighter than $B$\,=\,15.3 mag and have spectral types earlier than M\@.
\item All candidates identified in the successive GCS releases by
\citet{lodieu06}, \citet{lodieu07a}, \citet{dawson11}, and
\citet{dawson12} are recovered by our analysis but the assessment
of their membership changes slightly following the improvement on
the proper motions from the 2MASS/GCS cross-match to the two GCS epochs.
They are covered by the GCS DR10 $ZYJHK-$PM and SV samples.
\item The full catalogue of USco members published by \citet{luhman12c}
contains a total of 863 sources, including 381 brighter than 
$J$\,=\,11.5 mag, which are saturated on the GCS images.
Hence, our recovery rate of 405 sources out of (863$-$381)\,=\,482 in
the GCS is over 80\%. Similarly, we recovered 472 sources among the
806 with $HK$ photometry, which is over 98\% completeness because most
of the other objects are saturated in the GCS\@.
\item The recovery of candidates published by the remaining studies
is mainly biased due to the lack of overlap between those surveys
(red asterisks in Fig.\ \ref{fig_USco:coverage_DR10}) and the $ZYJHK-$PM
and GCS SV\@. Table \ref{tab_USco:early_summary} {\bf{demonstrates}} that
most of the sources published by previous studies are part of the 
region covered in $H,K$. The most incomplete recoveries are due to
the bright early-type members in the catalogues of \citet{preibisch98},
\citet{preibisch01}, and \citet{luhman12c} as discussed in the previous
bullets.
\end{itemize}
%

%
%
\begin{table}
  \centering
  \caption{Numbers of USco member candidates recovered in the full GCS
database using a matching radius of 3$''$ before running our SQL 
queries (GCS), in the $ZYJHK-$PM sample 
(DR10), in the SV area (SV), and in the $HK-$only region.
Papers dedicated to USco are listed below and ordered by year.
References are: \citet[][X-ray and proper motion samples]{preibisch98}, 
\citet{preibisch01}, \citet{preibisch02}, \citet{ardila00}, 
\citet{martin04}, \citet{slesnick06}, \citet{lodieu06}, \citet{lodieu07a}, 
\citet{lodieu08a}, \citet{dawson11}, \citet{lodieu11a}, \citet{dawson12},
\citet*{luhman12c}. The last column lists the percentage of sources
in the original paper recovered in the $ZYJHK-$PM coverage with and
without extinction.
}
  \label{tab_USco:early_summary}
  \begin{tabular}{@{\hspace{0mm}}l c c c c c@{\hspace{0mm}}}
  \hline
Survey reference    &  GCS      & DR10    & SV    & $HK$ & \%   \cr
  \hline
Preibisch1998\_Xray &   20/78   &    0/10 &    0/10  &  10/55  &  25.6  \cr
Preibisch1998\_PM   &  21/115   &    2/20 &    0/1   &  14/89  &  18.3  \cr
Preibisch2001       &  0/100    &    0/0  &    0/0   &   7/100 &   0.0  \cr
Preibisch2002       &  0/166    &    0/0  &    0/0   &  98/166 &   0.0  \cr
Ardila2000          &   17/20   &    1/2  &    0/0   &  14/20  &  85.0  \cr
Martin2004          &   10/40   &    2/6  &    2/4   &  36/36  &  25.0  \cr
Slesnick2006        &   7/38    &    2/2  &    2/5   &  33/34  &  18.4  \cr
Lodieu2006          &   15/15   &   15/15 &    0/0   &  15/15  & 100.0  \cr
Lodieu2007          &  129/129  &    0/0  &  64/129  &  97/129 & 100.0  \cr
Slesnick2008        &  38/145   &   10/12 &   16/26  & 127/140 &  26.2  \cr
Dawson2011          &   28/28   &   28/28 &    0/0   &  28/28  & 100.0  \cr
Dawson2012          &  116/116  &   74/74 &   35/42  & 116/116 & 100.0  \cr
Luhman2012          &  405/863  &   47/92 &  65/173  & 472/806 &  46.9  \cr
 \hline
Total               & 186/1566  & 125/158 &  73/202  & 651/1066 &  11.9  \cr
 \hline
\end{tabular}
\end{table}
%

%
%

%
%
\begin{figure*}
   \centering
   \includegraphics[width=0.495\linewidth]{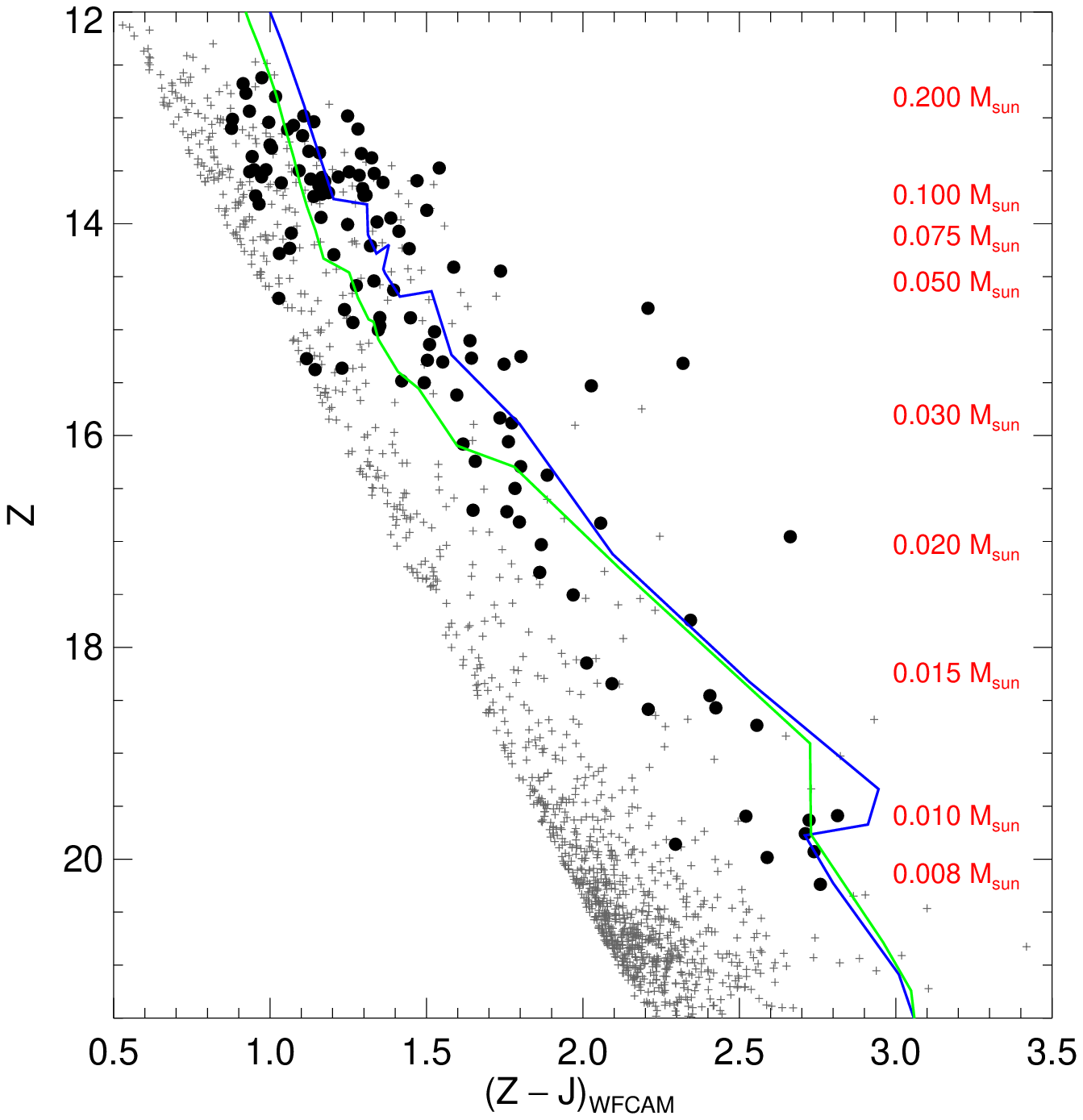}
   \includegraphics[width=0.495\linewidth]{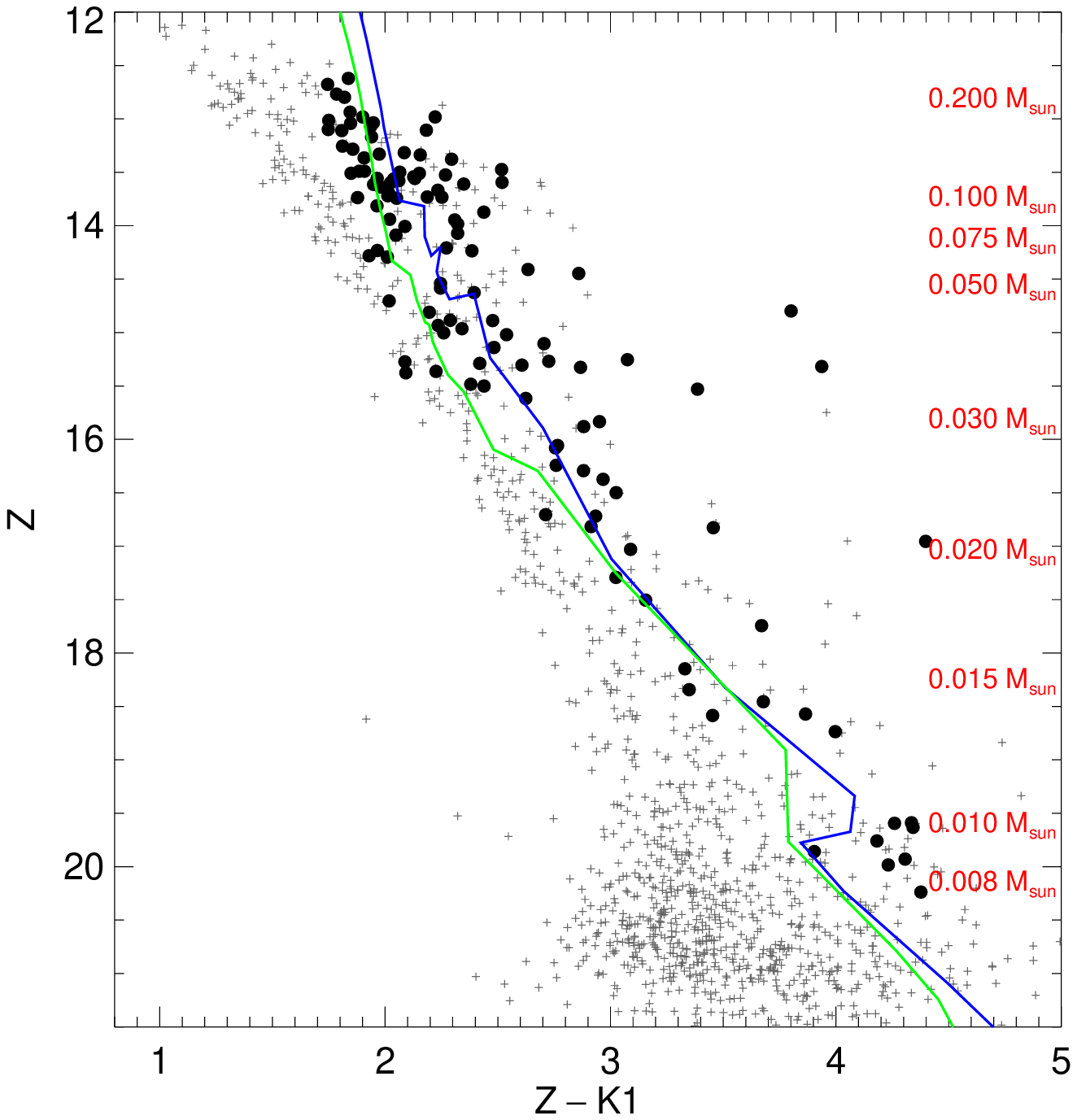}
   \includegraphics[width=0.495\linewidth]{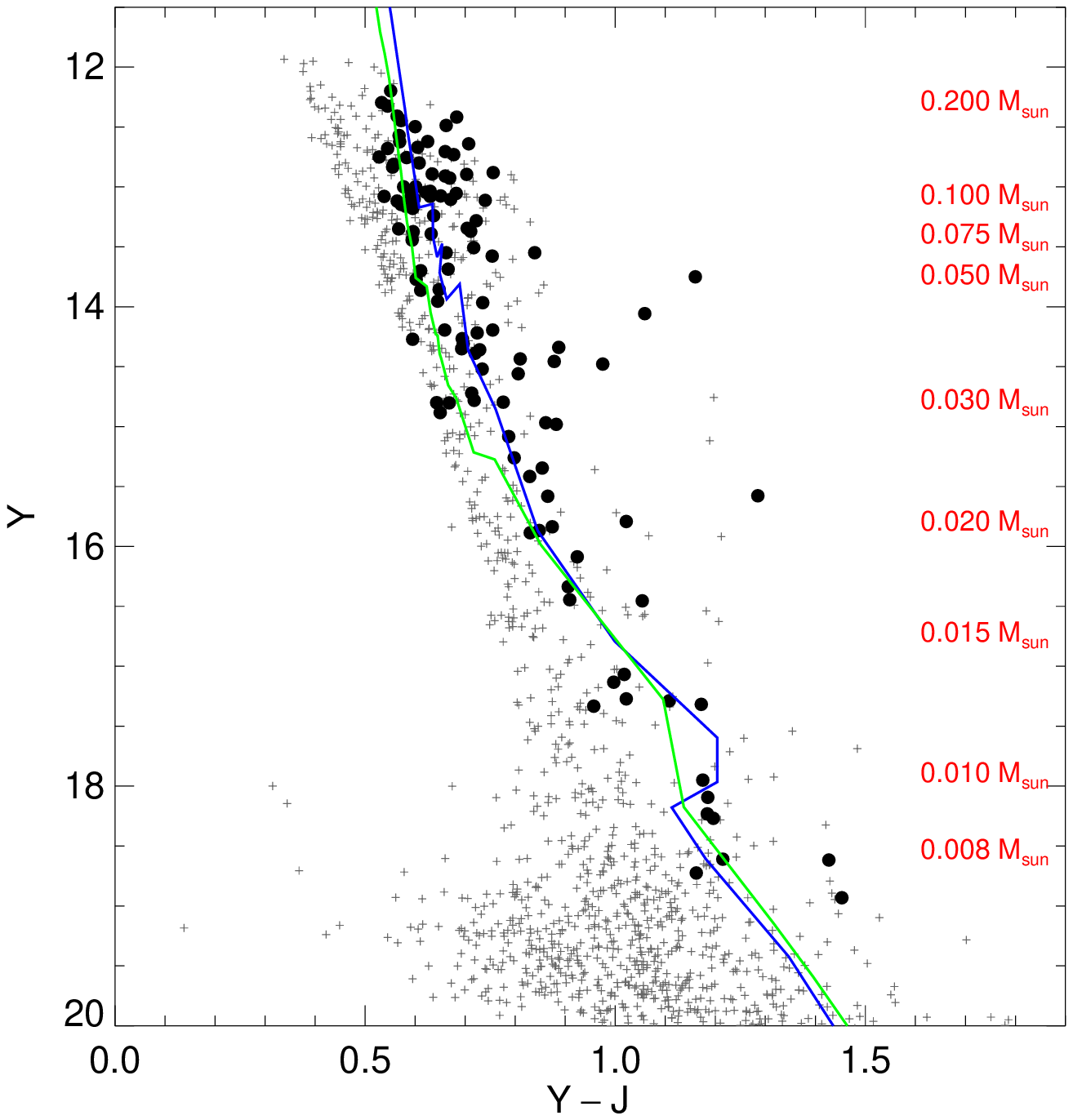}
   \includegraphics[width=0.495\linewidth]{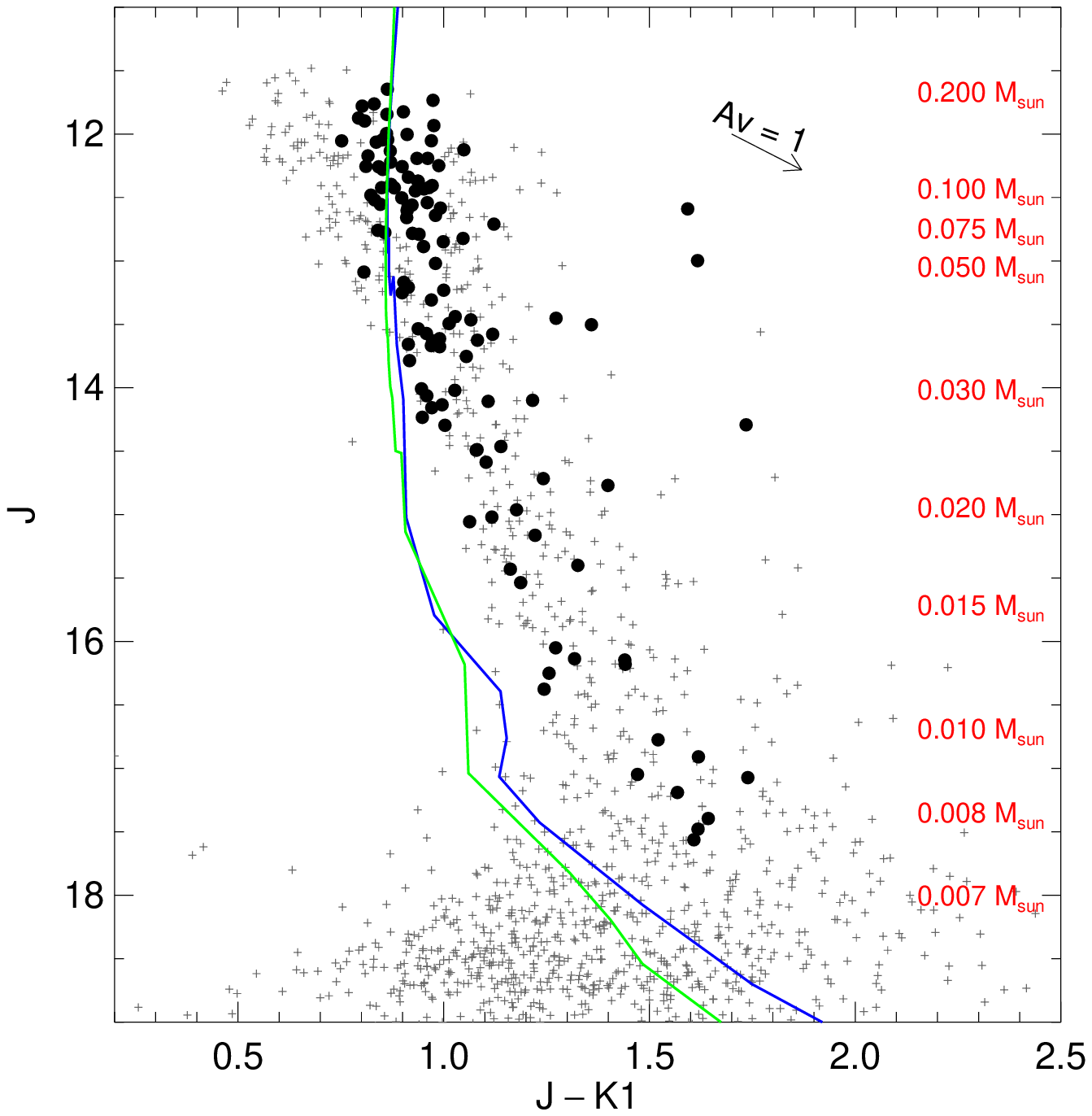}
   \caption{Same as figure \ref{fig_USco:YJK_cmds} but for the
region in USco affected by reddening.
}
   \label{fig_USco:YJK_cmds_Extinction}
\end{figure*}
%

%
%
\begin{figure*}
   \centering
   \includegraphics[width=0.495\linewidth]{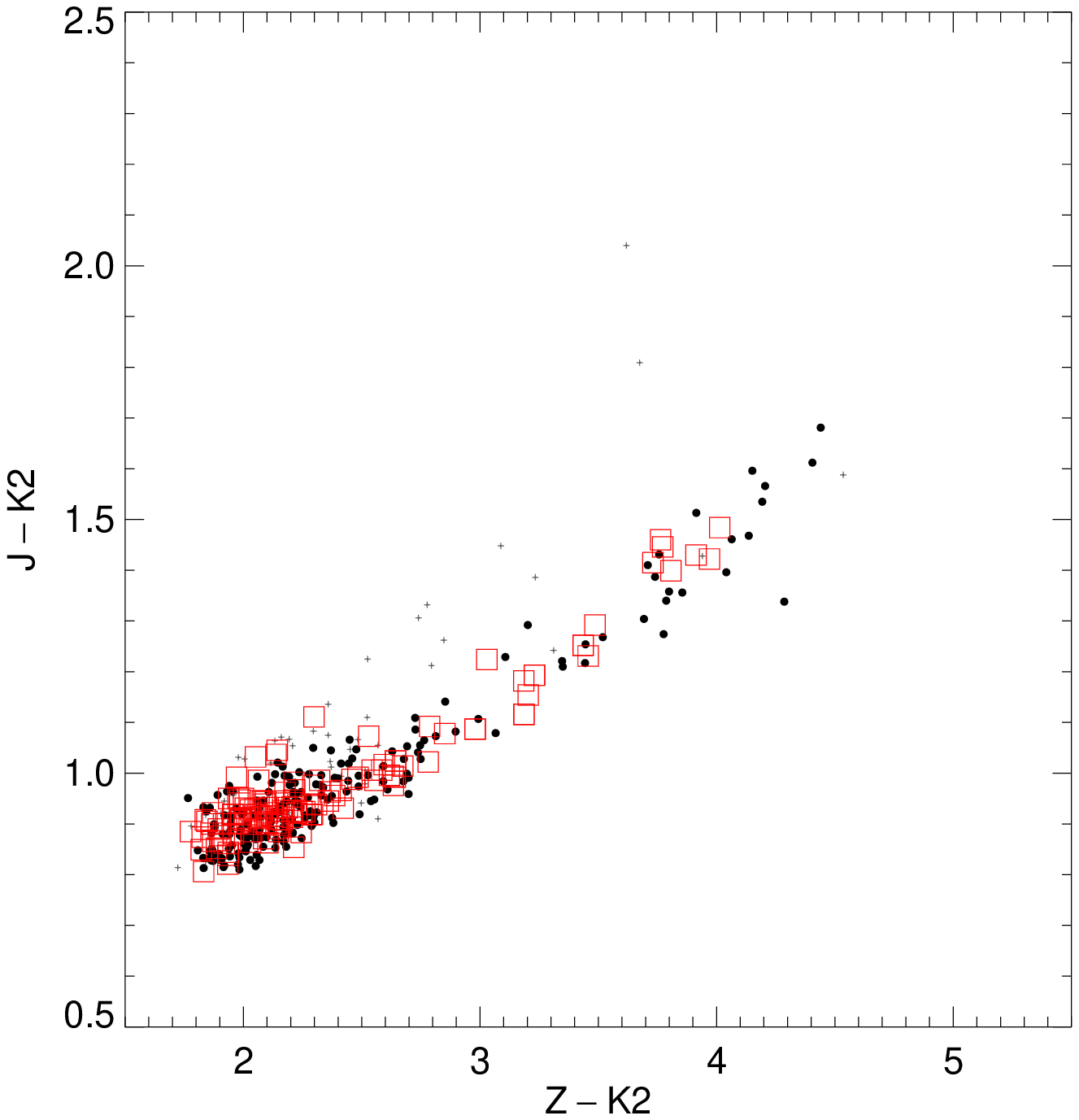}
   \includegraphics[width=0.495\linewidth]{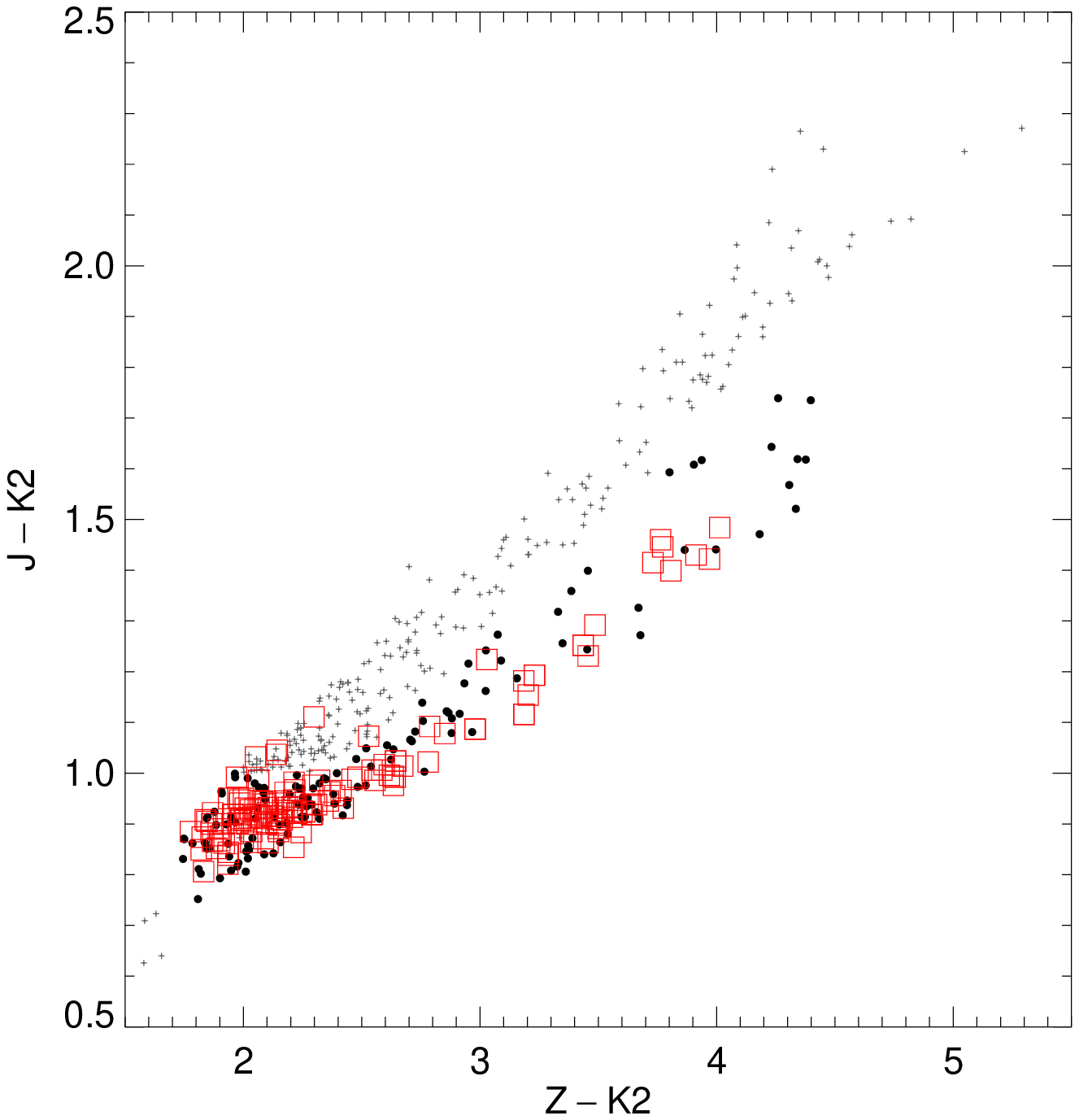}
   \caption{($Z-K$,$J-K$) colour-colour diagram for the photometric
and astrometric candidates (grey crosses) in the region free of extinction 
(left) and the region affected by reddening (right) of the $ZYJHK-$PM
sample. Overplotted as red open squares are known spectroscopic members.
USco candidate members identified in our study are plotted as black dots.
}
   \label{fig_USco:ZKJK_ccd}
\end{figure*}
%

%
%
%
\section{New USco low-mass and brown dwarf member candidates}
\label{USco:new_cand}

\subsection{Astrometric selection}
\label{USco:new_cand_PM}

After the original photometric selection and the recovery of known member 
candidates, we plotted them as red open squares in 
Fig.\ \ref{fig_USco:diagram_VPD}. We observe two groups of objects, one 
centered on (0,0) made of field objects, and another one depicting the 
position of USco. We measured mean (relative) proper motions of 
$-$8.6 and $-$19.6 mas/yr in right ascension and declination, respectively, 
compared to the absolute values of $-$11 and $-$25 mas/yr from Hipparcos 
\citep{deBruijne97,deZeeuw99}. We noticed the same effect in the 
Pleiades \citep{lodieu12a}, 
$\alpha$\,Per \citep{lodieu12c}, and Praesepe \citep{boudreault12} 
because the GCS provides {\it{relative}} motions rather than 
{\it{absolute}} motions as in the case of Hipparcos.
We applied a 3$\sigma$ astrometric selection using the error bars 
for each source from the GCS in both directions, leaving 87 of 
the 186 known member candidates in the $ZYJHK-$PM sample.

We applied the same 3$\sigma$ astrometric selection to the full sample 
of point sources towards USco, both for the sample free of extinction
and the one affected by reddening. In the former case, we are left with 
700 of the original 9351 sources, and, in the latter, 1357 of the 20,031
objects (grey crosses in Fig.\ \ref{fig_USco:diagram_VPD}).
We plot these proper motion member candidates as grey crosses in the 
colour-magnitude diagrams displayed in Fig.\ \ref{fig_USco:YJK_cmds}.
We tested the influence of our choice of the relative proper motion
values in right ascension and declination by adding and removing
1 mas/yr in both directions (about 20\% of the mean error bars on the
proper motions). {\bf{We}} found that the numbers of candidates
would change by less than 5.4\% (677 and 738 candidates in the worst cases
compared to 700) and 8.3\% (1253 and 1468 compared to 1357) in the case 
of the $ZYJHK$-PM samples without and with reddening, respectively.

For the remaining areas of the association, we measured the proper 
motions from the 2MASS/GCS cross-match, whose accuracy is about twice
worse than the GCS proper motions (10 mas/yr down to $J$\,=\,15.5 mag). 
The mean proper motion of known spectroscopic members in the 
SV area is $-$8.5 and $-$19.2 mas/yr in right ascension and declination, 
respectively. As described above, those values differ from the absolute 
mean proper motion of USco but we used them for our 2MASS/GCS 
astrometric selection. We note that these values are very similar
to the mean proper motions derived from the two GCS epochs.
We are left with 242 sources out of the 430 
original photometric candidates in the GCS SV area, after applying
a 2$\sigma$ selection (Table \ref{tab_USco:candidates_GCSDR10_SV}).
Changing the mean values of the proper motions in each direction
by $\pm$1 mas/yr results in a number of member candidates that differs
by less than 3.75\%.

For the $HK$-only area, the situation is worse than for the SV region 
because we have only two bands available ($H+K$) where the cluster sequence 
is not so well separated from the field stars along the line of the 
association as in the ($Z-J$,$Z$) colour-magnitude diagram. Hence, our 
final $HK$ sample will be significantly more contaminated than the 
aforementioned samples. First, we applied a conservative photometric 
selection in the ($H-K$,$H$) diagram by considering only point sources 
to the right of a line running from ($H-K$,$H$)\,=\,(0.2,12) to (0.7,18). 
After this first step, we are left with 63,520 candidates. Second, we 
applied a 2$\sigma$ astrometric selection (i.e.\ 2$\times$10 mas/yr or 
95.4\% completeness) in the $H$\,=\,12.5--15 
mag interval (corresponding to masses ranging from 0.12--0.175 M$_{\odot}$ 
to 0.015--0.02 M$_{\odot}$ for ages of 5 and 10 Myr, respectively),
yielding 976 candidates. Third, we applied a stricter photometric
selection in the ($H-K$,$H$) diagram, keeping sources to the right
of a line running from (0.31,12.5) to (0.7,16.5). This line was chosen
to recover all photometric and astrometric candidates from the SV-only
and $ZYJHK$-PM samples. We are left with 286 candidates in the $HK$ region 
(Fig.\ \ref{fig_USco:coverage_DR10}; 
Table \ref{tab_USco:candidates_GCSDR10_HKonly}).
We tested the influence of our astrometric selection by changing the
mean proper motion in RA and dec by $\pm$1 mas/yr, yielding in the
extreme cases 281 and 304 candidates i.e.\ a difference of 6.3\% in the
worst case compared to our original choice. The main uncertainty on the
number of candidates in the $HK$-only sample rather comes from the choice 
of the sigma in the astrometric selection: choosing 2.5$\sigma$ (99\%
completeness) and 3$\sigma$ (99.9\% completeness) would lead to 360 and 
412 candidates, respectively.

\subsection{Photometric selection}
\label{USco:new_cand_phot}

To further refine our list of USco member candidates we applied
additional photometric cuts in various colour-magnitude diagrams
for the $ZYJHK-$PM sample, defined as follows:
\begin{itemize}
\item ($Z-K$,$Z$) = (1.60,12.0) to (2.20,16.0)
\item ($Z-K$,$Z$) = (2.20,16.0) to (4.00,20.0)
\item ($Y-J$,$Y$) = (0.40,12.0) to (0.65,15.0)
\item ($Y-J$,$Y$) = (0.65,15.0) to (1.00,18.0)
\item ($J-K$,$J$) = (0.80,11.5) to (0.80,14.0)
\item ($J-K$,$J$) = (0.80,14.0) to (1.40,18.0)
\end{itemize}

The numbers of candidates afer the $Z-K$, $Y-J$, and $J-K$ photometric 
selections are 279, 257, and 252, respectively, demonstrating that the
most influencial criteria are the astrometric and $Z-K$ selections.
Indeed, the additional {\bf{$Y-J$}} and $J-K$ selection remove small numbers
of cluster member candidates.
We stress that these cuts were chosen to recover known spectroscopic
members from earlier surveys (see Section \ref{USco:status_old_cand}).
These photometric selections returned 252 candidates in the region
free of extinction (filled black dots in Fig.\ \ref{fig_USco:YJK_cmds})
and 396 in the region affected by reddening 
(filled black dots in Fig.\ \ref{fig_USco:YJK_cmds_Extinction}). 
Similarly, we identified 84 member candidates in the SV region of 
the GCS\@.

We tested the influence of the choice of our selection lines on
the final numbers of candidates, in the specific case of the $ZYJHK$-PM
sample without extinction. We shifted each selection line to the left
and to the right by 0.1 mag, which corresponds roughly to the error
on the colour (i.e.\ 1$\sigma$) at the faint end of the sequence at 
$J$\,=\,18 mag. This shift corresponds to 2.5$\sigma$ at 
$J$\,=\,17 mag. We found that the shift to the blue of the six selection 
lines enumerated above yields roughly 20--30\% more cluster member
candidates. Similarly, a shift to the red gives about 20\% less candidates
in the ($Z-K$,$Z$) diagram and 41--43\% less in the other two diagrams.

We know that the level of contamination will be high in the region
affected by reddening. Hence, we applied an additional photometric
criterion in the ($Z-K$,$J-K$) two-colour diagram 
(Fig.\ \ref{fig_USco:ZKJK_ccd}) to remove giants and reddened stars 
based on the location of previously-known spectroscopic members 
\citep[red open squares;][]{lodieu11a}. We selected sources satisfying
the criteria:
\begin{itemize}
\item ($J-K$)\,$\geq$\,1.0 for ($Z-K$) between 1.7 and 2.4
\item Sources below the line defined by ($Z-K$,$J-K$) = (2.4,1.0) and (4.4,1.85)
\end{itemize}

This selection returned 201 and 120 sources (corresponding to 80\%
and 30\% of the candidates left after the astrometric and the first three
photometric selections) in the region with and without reddening, 
respectively. Only two candidates (or 2.5\%) were rejected
in the SV sample after applying this additional {\bf{criterion}}.
We list the coordinates, photometry, and proper motions of the 
candidates identified in $ZYJHK-$PM (including known members published 
by other groups) in Tables \ref{tab_USco:candidates_GCSDR10_noExt} 
and \ref{tab_USco:candidates_GCSDR10_Extinction} in the Appendix for 
the regions with and without reddening, respectively.
We display their distribution in Fig.\ \ref{fig_USco:coverage_DR10}.
We provide the list of member candidates within the SV area in 
Table \ref{tab_USco:candidates_GCSDR10_SV}.

We tested the influence of the choice of the crude selection in
the ($Z-J$,$Z$) colour-magnitude diagram by choosing a line {\bf{shifted}} 
by 0.1 mag to the left of the original choice. We applied again 
the same aforementionned criteria and arrived at the same numbers of
candidates for the two $ZYJHK-$PM samples and the SV-only sample.

\subsection{Catalogue summary}
\label{USco:new_cand_phot_summary}

To summarise, we have identified a total of 688 sources in four different 
regions within USco: 195 and 111 in the $ZYJHK-$PM regions without 
reddening and with extinction, 79 in the SV area, and 276 in the $HK$-only 
region. We found 11 sources in common among them, leaving 677 USco member 
candidates. We cross-matched this list with itself and found 4, 12, and 
15 sources within 10, 50, and 100 arcsec of each other, pointing towards 
binary fractions for wide common proper motions of 0.6\%, 1.8\%, and 
2.2\% for projected separations of 1450 au, 7250 au, and 14500 au, 
respectively.

%
%
\section{Variability at young ages}
\label{USco:variability}

\begin{figure}
   \includegraphics[width=\linewidth]{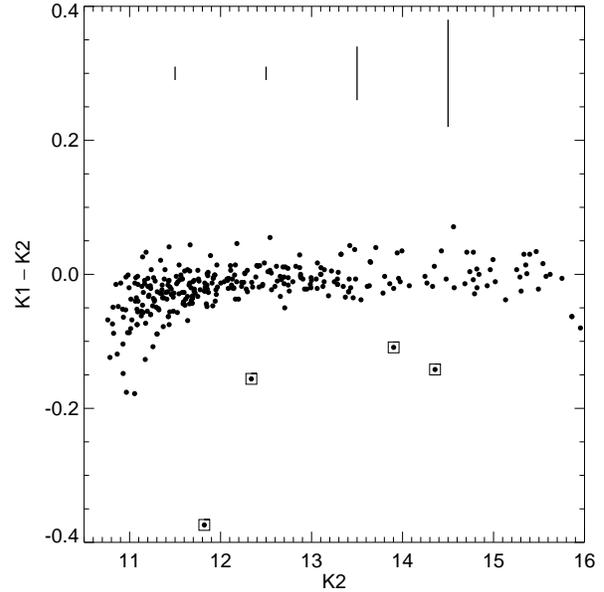}
   \caption{Difference in the $K$ magnitude ($K$1--$K$2) as a function 
of the $K$2 magnitude for all USco member candidates with proper 
motions from GCS DR10\@. The four potential variable sources are
highlighted with large open squares.
Typical error bars on the colour shown as vertical dotted lines are added
at the top of the plot.
}
   \label{fig_USco:diagram_VAR}
\end{figure}

We investigate the variability of low-mass stars and brown dwarfs 
in USco using the two $K$-band epochs provided by the GCS\@. 
Figure \ref{fig_USco:diagram_VAR} shows the ($K$1--$K$2) vs $K$2
diagram for USco member candidates in the $ZYJHK-$PM sample.
This analysis is not possible for the SV sample because no second 
$K$-band epoch is available.

The brightening in the $K$1 = 10.5--11.5 mag range is due to the 
difference in saturation between the first and second 
epoch, of the order of 0.5 mag both in the saturation and completeness 
limit. This is understandable because the exposure times have been 
doubled for the second epoch with relaxed constraints on the seeing 
requirement and weather conditions. We excluded those objects from our 
variability study. Overall, the sequence indicates consistent photometry
between the two $K$ epochs with very few objects being variable.

At first glance, we spotted four potential variables in 
Figure \ref{fig_USco:diagram_VAR} (Table \ref{tab_USco:VAR_cand}). 
The GCS images do not show anything anomalous so we did {\bf{not}} proceed 
further. We selected variable objects by looking at the standard deviation,
defined as 1.48$\times$ the median absolute deviation which is the 
median of the sorted set of absolute values of deviation from the 
central value of the $K1-K2$ colour. We identified one potential 
variable object with a difference of 0.37 mag in the 11.5--12 mag
range whereas the other three fainter candidates with differences 
between 0.1 and 0.16 mag lie just below the 3$\sigma$ of the median
absolute deviations of 0.05--0.06 mag. These small variations of the
order of 0.1--0.15 mag can be interpreted by the presence of cool
spots in low-mass stars \citep[e.g.][]{scholz09b}.

We conclude that the level of $K$-band variability at 5--10 Myr is small, 
with standard deviations of the order of 0.06 mag range, suggesting that it
cannot account for the dispersion in the cluster sequence. We arrived at
the same conclusions in the case of the Pleiades \citep{lodieu12a},
$\alpha$\,Per \citep{lodieu12c}, and Praesepe \citep{boudreault12}
although these clusters are older (85, 120, and 590 Myr, respectively).

%
%
%

%
%
\begin{table}
  \caption{Potential variable candidates in our USco $ZYJHK-$PM 
  sample of low-mass stars and brown dwarfs.}
  \centering
  \begin{tabular}{c c c c c}
  \hline
R.A.      &  dec      & $K1$    &  $K2$  & $\Delta K$  \cr
  \hline
15:52:33.93 & $-$26:51:12.4  & 11.447 & 11.821 & $-$0.374 \cr
16:23:23.06 & $-$29:01:33.4  & 12.182 & 12.338 & $-$0.156 \cr
15:47:22.82 & $-$21:39:14.3  & 14.215 & 14.357 & $-$0.142 \cr
16:23:22.02 & $-$26:09:55.6  & 13.793 & 13.902 & $-$0.109 \cr
 \hline
  \label{tab_USco:VAR_cand}
\end{tabular}
\end{table}
%

%
%
\begin{figure*}
   \centering
   \includegraphics[width=0.49\linewidth]{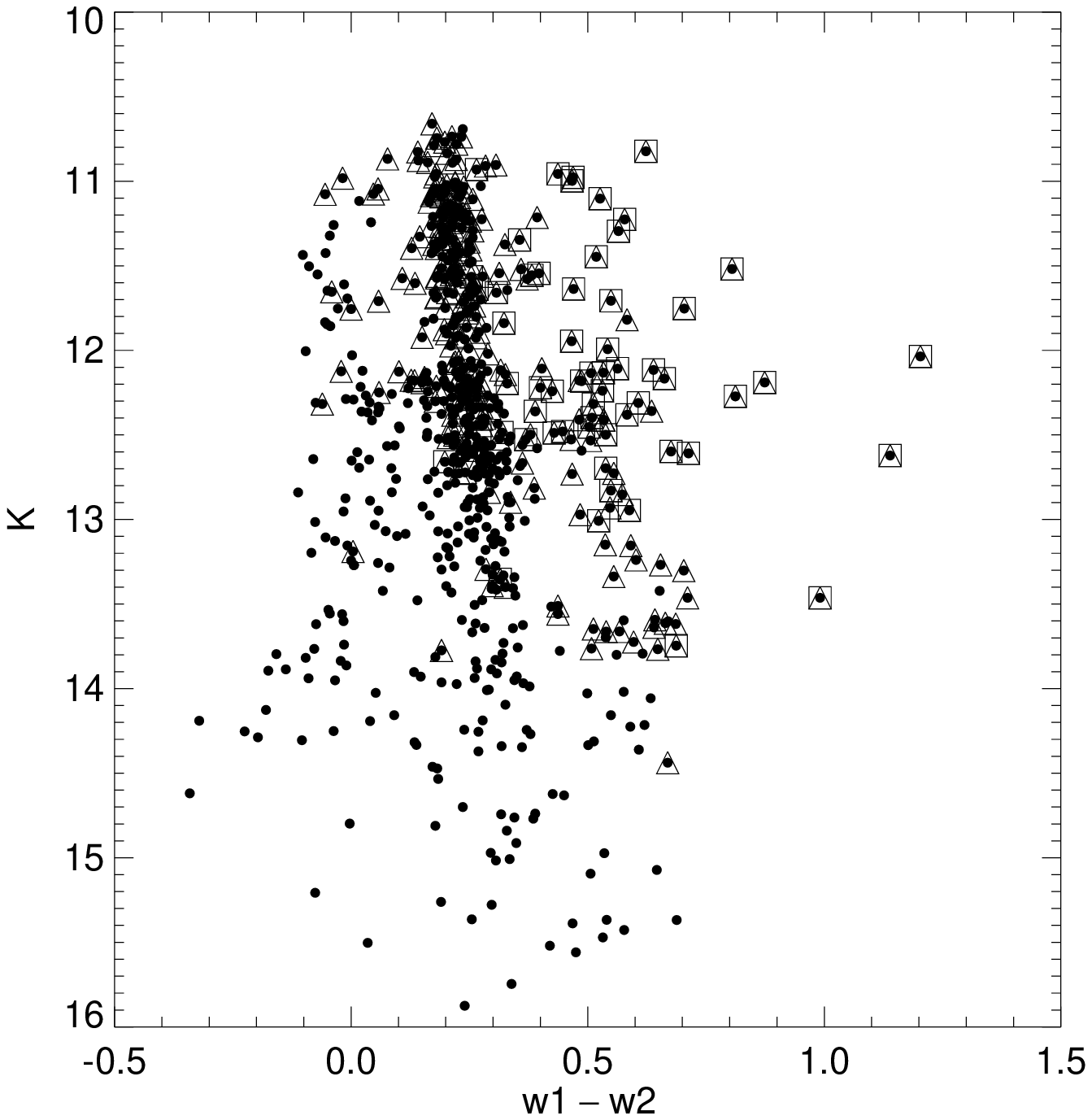}
   \includegraphics[width=0.49\linewidth]{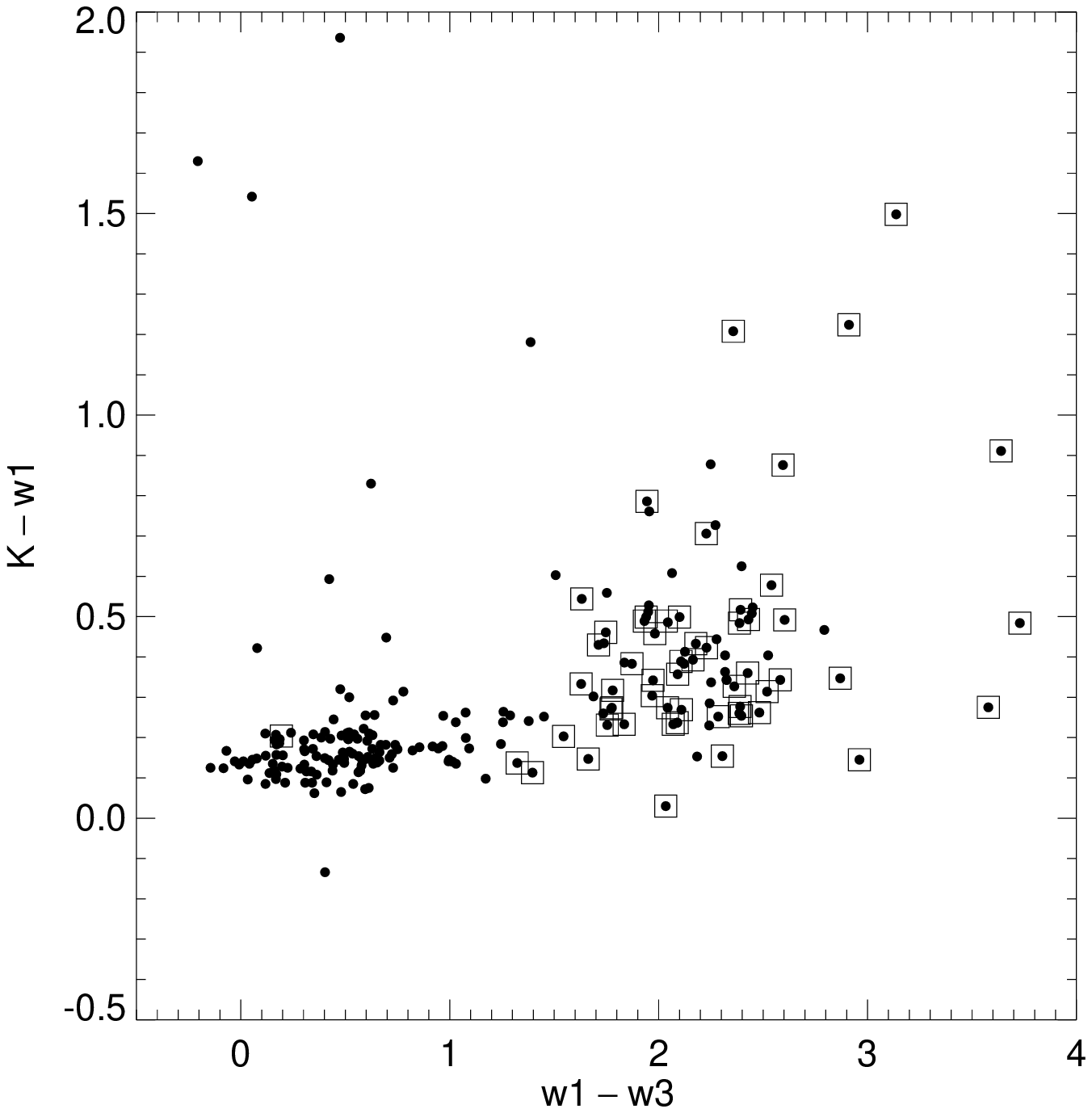}
   \caption{($w1-w2$,$K$) and ($w1-w3$,$K-w1$) diagrams for USco members
with WISE counterparts (black dots). Open triangles and open squares are 
sources detected in $w3$ and $w3+w4$, respectively.
}
   \label{fig_USco:diagram_WISE}
\end{figure*}
%

%
%
\section{Disk frequency in USco}
\label{USco:disk_freq_USco}

We cross-correlated our lists of candidates with the Wide Field Infrared 
Survey Explorer (WISE) All-sky release \citep{wright10} using a matching 
radius of three arcsec. We found a total of 660 matches, divided into 
194 in the $ZYJHK-$PM, 111 in the $ZYJHK-$PM region with extinction,
79 objects in the SV-only area, and 276 counterparts in the $HK$-only 
region, respectively (Table \ref{tab_USco:candidates_GCSDR10_WISE}). 
We looked at the WISE images and found that all WISE counterparts to 
the GCS objects are detected in $w1$ and $w2$. We classified the sources 
listed in Table \ref{tab_USco:candidates_GCSDR10_WISE} in two categories: 
224 (33.9\%) {\bf{source}} detections in $w3$ but not in 
$w4$ and 58 (8.8\%) objects detected in $w3$ and $w4$ marked as 1110 
(open triangles in Figure \ref{fig_USco:diagram_WISE}) and 1111 
(open squares in Figure \ref{fig_USco:diagram_WISE}), respectively.
The latter objects detected in all WISE bands are unambiguous 
disk-bearing low-mass stars and brown dwarfs. 

We considered two different methods to estimate the disk frequency of 
USco members with masses below 0.2 M$_{\odot}$, according to the
BT-Settl models \citep{allard12}. We plot in 
Figure \ref{fig_USco:diagram_WISE} the ($w1-w2$,$K$) and ($K-w1$,$w1-w3$)
diagrams. The former represents a good discriminant to separate disk-less
and disk-bearing objects according to \citet{dawson12}. We note that
we chose $K$ as the infrared band rather than $J$ as in \citet{dawson12}
because all targets in our four samples have $K$-band photometry.
This diagram is efficient in Taurus but may not be the best criterion
for USco where many transition disks are found \citep{riaz12a}.
The ($K-w1$,$w1-w3$) colour-colour diagram, however, clearly separates
brown dwarf and M dwarf disks \citep{penya12a} as well as primordial 
disks in USco \citep[Figure 5 of][]{riaz12a}.

We observe three sequences in the ($w1-w2$,$K$) diagram
(left-hand side panel of Figure \ref{fig_USco:diagram_WISE}): one sequence 
to the left likely made of non-members mainly from the $ZYJHK-$PM region 
with extinction and the $HK$-only area where our contamination is expected 
to be higher than in the other two areas ($ZYJHK-$PM without reddening 
and SV-only). Optical spectroscopy is needed to confirm our claims though. 
Broadly, we have the same number of sources with disks (16\% or 107 
sources with $w1-w2$\,$\geq$\,0.4 mag) as potential contaminants 
(100 sources or 15\% wit $w1-w2$\,$\leq$\,0.1 mag), implying that the 
frequency if disk-bearing USco members lies between 107/600\,=\,16.2\% and 
107/(660$-$100)\,=\,19.1\%, following the arguments of \citet{dawson12}. 
However, these numbers are lower limits because 17/58 (29.3\%) $w3+w4$ 
detections have $w1+w2$ colours bluer than 0.4 mag i.e.\ lie in the 
middle sequence, yielding corrected disk fractions of 18.8 and 22.1\%. 
We are certainly missing some disk-bearing
members hiding among $w3$ detections but it is harder to quantify without 
a cautious fitting of the spectral energy distributions (beyond the scope 
of this paper), as discussed in details in \citet{riaz09b} and 
\citet{dawson12}. Nonetheless, we can place an upper limit of 
(224$+$58)/660\,=\,42.7\% on the overall disk fraction for USco low-mass 
stars and brown dwarfs. Similarly, we can set a lower limit of 
6/58\,=\,10\% based on the six unambiguous disk objects with 
$w1+w2$\,$\geq$\,0.8 among sources detected in all four WISE bands.

We observe two groups of objects in the ($K-w1$,$w1-w3$) diagram
(right-hand side panel of Figure \ref{fig_USco:diagram_WISE}) depicting 
photospheric sources ($w1-w3$\,$\leq$\,1.2 mag) and disk-bearing 
members ($w1-w3$\,$\geq$\,1.5 mag). Objects in the middle may be good 
candidates to transition disks \citep{riaz12a}. Among the 282
USco member candidates detected in three or four WISE bands, we
have 89 sources with $w1-w3$\,$\geq$\,1.5 mag, implying a disk
fraction of 31.5\% (most likely range of 26.6--37.1\%). This fraction
would increase by 5\% if we include the potential transition disks.

The disk frequencies derived by both methods are consistent within the
error bars although the second one is on average higher. Our values
should be compared with the 23$\pm$5\% of \citet{dawson12} for USco 
brown dwarfs and 25$\pm$3\% of \citet{luhman12c} for M4--L2 members. 
For earlier-type members the disk {\bf{fractions}} range from 10\% for K0--M0 
dwarfs \citep{luhman12c} to 19\% for K0--M5 dwarfs \citep{carpenter06}.
Our disk fraction for USco substellar members is consistent with the
42$\pm$12\% and 36$\pm$8\% disk frequencies for brown dwarfs in 
IC\,348 \citep[1--3 Myr;][]{luhman05a} and $\sigma$\,Orionis 
\citep[1--8 Myr;][]{penya12a}.

%
%
\section{The Initial mass function}
\label{USco:IMF}

We derive the cluster luminosity and system mass functions from our 
astrometric and photometric sample of $\sim$320 USco member candidates 
distributed over $\sim$50 square degrees (the $ZYJHK-$PM sample). 
We did not attempt to correct the mass function for binaries for two
reasons. First, the presence of disks around USco low-mass stars and
brown dwarfs tend to displace these sources to the right-hand side of
the sequence, implying that disentangling binaries from disk-bearing
members is harder than in the case of more mature clusters like the
Pleiades \citep{lodieu12a} and Praesepe \citep{boudreault12}.
Second, a recent high-resolution imaging survey of 20 USco spectroscopic
members fainter than $J$\,=\,15 mag by \citet{biller11} resolved only 
one binary, pointing towards a binary frequency lower than 10\%. This
fraction is lower than the uncertainty on the number of sources per
mass bin at the low-mass end, assuming Gehrels error bars.

\subsection{The age of USco }
\label{USco:IMF_age}

\citet{deZeeuw85} and \citet{deGeus89} 
derived an age of 5--6 Myr comparing sets of photometric data with
theoretical isochrones available at that time. Later, \citet{preibisch99}
confirmed that USco is likely 5 Myr-old with a small dispersion on the 
age by placing about 100 spectroscopic members on the 
Hertzsprung-Russel diagram. This age estimate was later supported
by \citet{slesnick08} from a wide-field optical study of low-mass
stars and brown dwarfs with spectroscopic membership: these authors
derived a mean age of 5 Myr with an uncertainty of 3 Myr taking into
account all possible sources of uncertainties. More recently,
\citet{pecaut12} argued for an age older (11$\pm$2 Myr) from
an analysis of the F-type members of the full Scorpius-Centaurus
association. They determined older ages for all three subgroups forming
Scorpius-Centaurus, including ages of 16 and 17 Myr for 
Upper Centaurus-Lupus and Lower Centaurus-Crux, which are older
than the ages derived by \citet{song12} from the abundances of lithium
in the spectra of F/K members of the association.

In this work, we adopt an age of 5 Myr for USco but we will also
investigate the influence of age on the shape of the mass function 
(Section \ref{USco:IMF_MF}).

%
%
\begin{figure*}
   \includegraphics[width=0.495\linewidth]{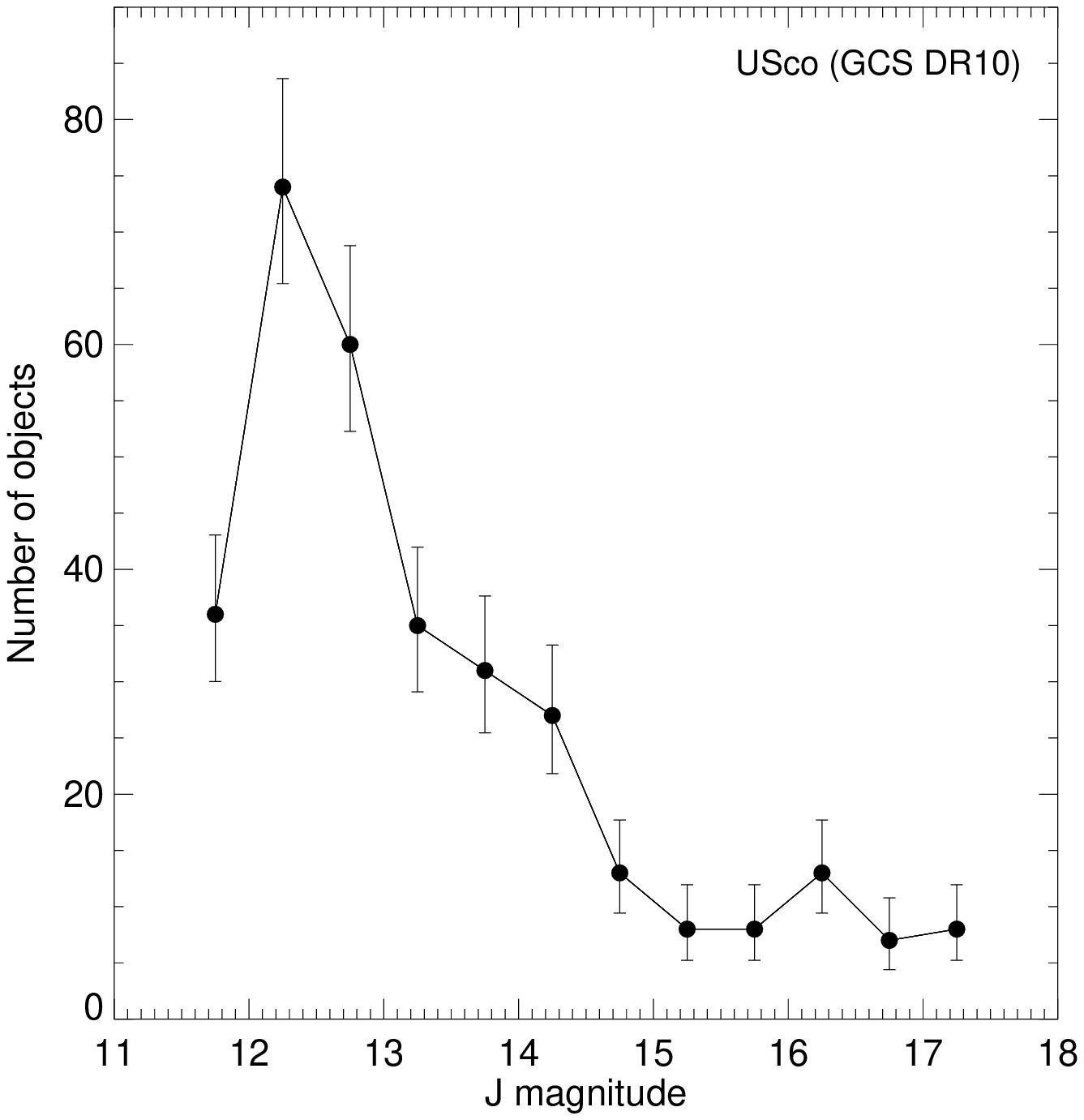}
   \includegraphics[width=0.495\linewidth]{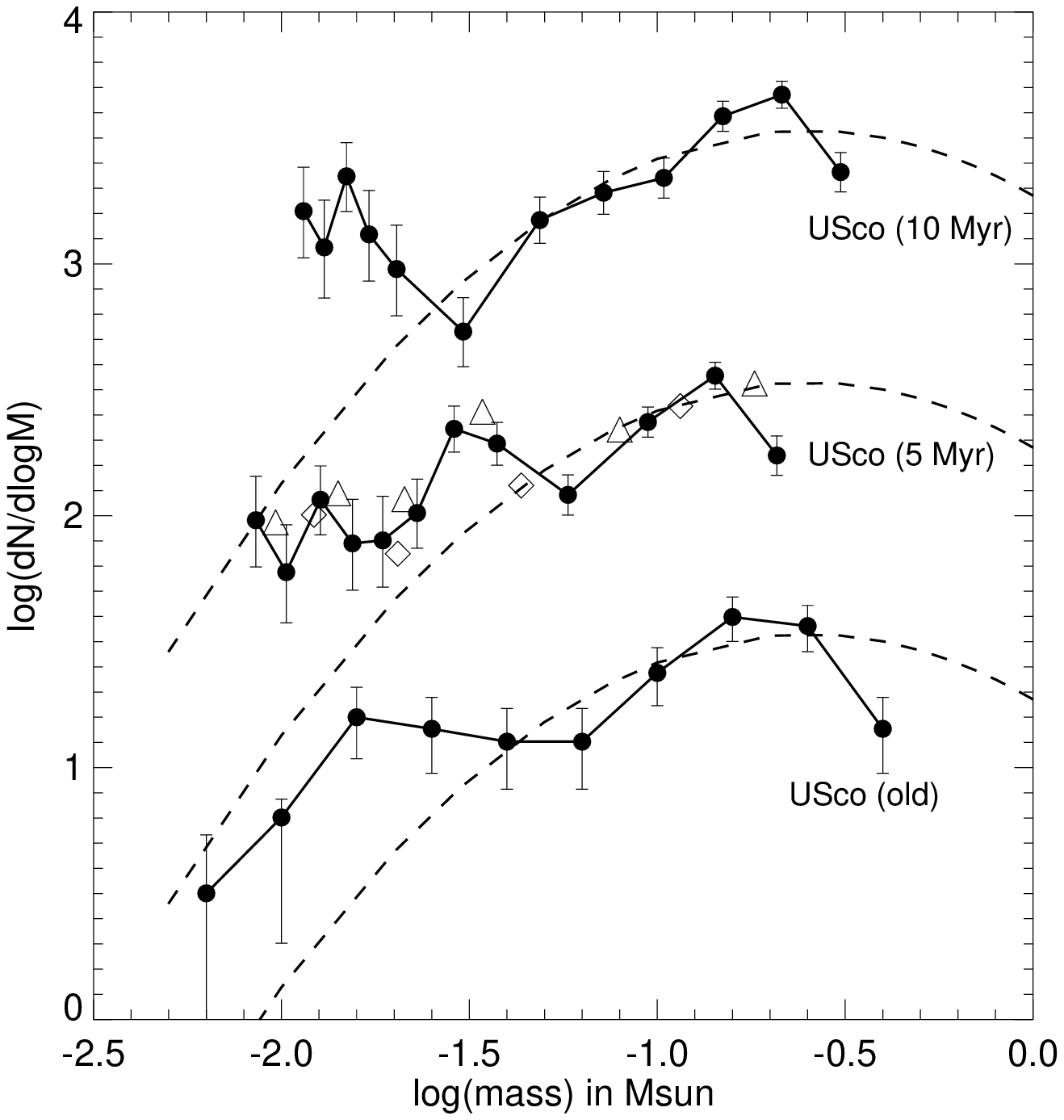}
   \caption{Luminosity (left) and mass (right) functions derived from our 
   $ZYJHK-$PM sample of USco member candidates.
We plot the mass function for 5 and 10 Myr, ages quoted in the literature
for USco as well as the mass function derived from the spectroscopic
sample of \citet{lodieu11a}. Overplotted as a dashed line is the 
log-normal field mass function \citep{chabrier05a}.
The binned mass functions are shown as open triangles (six bins) and
open diamonds (four bins).
} 
   \label{fig_USco:LF_and_MF}
\end{figure*}
\subsection{The cluster luminosity function}
\label{USco:IMF_LF}

We construct the cluster luminosity function from our astrometric and 
photometric $ZYJHK-$PM sample of 320 candidate members spanning 
$J$\,=\,11--17.5 mag. We display the luminosity function divided
into bins of 0.5 mag in the left panel of Fig.\ \ref{fig_USco:LF_and_MF}
with Gehrels error bars \citep{gehrels86}. We note that the first
and last bins are lower limits because of incompleteness at the bright
and faint ends. We observe that the number of objects {\bf{decreases}} with 
fainter magnitudes i.e.\ cooler temperatures.

%
%
\begin{table*}
  \caption{Values for the luminosity and mass functions for USco for an age
  of 5 Myr. We assumed a distance of 145 pc and employed the 
BT-Settl theoretical isochrones to transform magnitudes into masses
\citep{allard12}. 
The mass function is plotted in Fig.\ \ref{fig_USco:LF_and_MF}.
}
  \label{tab_USco:LF_MF_5Myr}
  \begin{tabular}{c c c c | c c c | c c c | c c c}
  \hline
Mag range &   Nb\_obj & Mass range  & Mid-mass & dN   & errH & errL & dlogM & errH & errL & dN/dM & errH & errL \cr
  \hline
11.5--12.0 &   36 & 0.2470--0.1690 & 0.2080 &  36.00 &   7.06 &   5.98 &  461.54 &   90.54 &   76.66 &  2.34 & 0.18 & 0.18 \\
12.0--12.5 &   74 & 0.1690--0.1160 & 0.1425 &  74.00 &   9.65 &   8.59 & 1396.23 &  182.00 &  162.03 &  2.66 & 0.12 & 0.12 \\
12.5--13.0 &   60 & 0.1160--0.0728 & 0.0944 &  60.00 &   8.79 &   7.73 & 1388.89 &  203.57 &  178.93 &  2.47 & 0.14 & 0.14 \\
13.0--13.5 &   35 & 0.0728--0.0429 & 0.0578 &  35.00 &   6.98 &   5.89 & 1170.57 &  233.42 &  197.15 &  2.18 & 0.18 & 0.18 \\
13.5--14.0 &   31 & 0.0429--0.0320 & 0.0374 &  31.00 &   6.63 &   5.55 & 2844.04 &  608.69 &  508.74 &  2.39 & 0.19 & 0.20 \\
14.0--14.5 &   27 & 0.0320--0.0256 & 0.0288 &  27.00 &   6.27 &   5.17 & 4218.75 &  979.35 &  808.13 &  2.45 & 0.21 & 0.21 \\
14.5--15.0 &   13 & 0.0256--0.0203 & 0.0230 &  13.00 &   4.71 &   3.57 & 2452.83 &  888.32 &  673.72 &  2.11 & 0.31 & 0.32 \\
15.0--15.5 &    8 & 0.0203--0.0169 & 0.0186 &   8.00 &   3.96 &   2.78 & 2352.94 & 1164.13 &  818.79 &  2.00 & 0.40 & 0.43 \\
15.5--16.0 &    8 & 0.0169--0.0140 & 0.0155 &   8.00 &   3.96 &   2.78 & 2758.62 & 1364.84 &  959.96 &  1.99 & 0.40 & 0.43 \\
16.0--16.5 &   13 & 0.0140--0.0114 & 0.0127 &  13.00 &   4.71 &   3.57 & 5000.00 & 1810.81 & 1373.35 &  2.16 & 0.31 & 0.32 \\
16.5--17.0 &    7 & 0.0114--0.0092 & 0.0103 &   7.00 &   3.78 &   2.60 & 3181.82 & 1719.95 & 1180.94 &  1.88 & 0.43 & 0.46 \\
17.0--17.5 &    8 & 0.0092--0.0079 & 0.0086 &   8.00 &   3.96 &   2.78 & 6153.85 & 3044.65 & 2141.45 &  2.08 & 0.40 & 0.43 \\
 \hline
\end{tabular}
\end{table*}
%

%
%
\begin{table*}
  \caption{Values for the luminosity and mass functions for USco for an age
  of 10 Myr. We assumed a distance of 145 pc and employed the 
BT-Settl theoretical isochrones to transform magnitudes into masses
\citep{allard12}. The mass function is plotted in 
Fig.\ \ref{fig_USco:LF_and_MF}. 
}
  \label{tab_USco:LF_MF_10Myr}
  \begin{tabular}{c c c c | c c c | c c c | c c c}
  \hline
Mag range &   Nb\_obj & Mass range  & Mid-mass & dN   & errH & errL & dlogM & errH & errL & dN/dM & errH & errL \cr
  \hline
11.5--12.0 &   36 & 0.3620--0.2530 & 0.3075 &  36.00 &   7.06 &   5.98 &  330.28 &   64.79 &   54.85 &  2.36 & 0.18 & 0.18 \\
12.0--12.5 &   74 & 0.2530--0.1760 & 0.2145 &  74.00 &   9.65 &   8.59 &  961.04 &  125.27 &  111.53 &  2.67 & 0.12 & 0.12 \\
12.5--13.0 &   60 & 0.1760--0.1230 & 0.1495 &  60.00 &   8.79 &   7.73 & 1132.08 &  165.93 &  145.85 &  2.59 & 0.14 & 0.14 \\
13.0--13.5 &   35 & 0.1230--0.0852 & 0.1041 &  35.00 &   6.98 &   5.89 &  925.93 &  184.63 &  155.95 &  2.34 & 0.18 & 0.18 \\
13.5--14.0 &   31 & 0.0852--0.0587 & 0.0720 &  31.00 &   6.63 &   5.55 & 1169.81 &  250.37 &  209.26 &  2.28 & 0.19 & 0.20 \\
14.0--14.5 &   27 & 0.0587--0.0387 & 0.0487 &  27.00 &   6.27 &   5.17 & 1350.00 &  313.39 &  258.60 &  2.17 & 0.21 & 0.21 \\
14.5--15.0 &   13 & 0.0387--0.0222 & 0.0305 &  13.00 &   4.71 &   3.57 &  787.88 &  285.34 &  216.41 &  1.73 & 0.31 & 0.32 \\
15.0--15.5 &    8 & 0.0222--0.0183 & 0.0203 &   8.00 &   3.96 &   2.78 & 2051.28 & 1014.88 &  713.82 &  1.98 & 0.40 & 0.43 \\
15.5--16.0 &    8 & 0.0183--0.0159 & 0.0171 &   8.00 &   3.96 &   2.78 & 3333.33 & 1649.18 & 1159.95 &  2.12 & 0.40 & 0.43 \\
16.0--16.5 &   13 & 0.0159--0.0139 & 0.0149 &  13.00 &   4.71 &   3.57 & 6500.00 & 2354.05 & 1785.36 &  2.35 & 0.31 & 0.32 \\
16.5--17.0 &    7 & 0.0139--0.0121 & 0.0130 &   7.00 &   3.78 &   2.60 & 3888.89 & 2102.16 & 1443.38 &  2.07 & 0.43 & 0.46 \\
17.0--17.5 &    8 & 0.0121--0.0108 & 0.0115 &   8.00 &   3.96 &   2.78 & 6153.85 & 3044.65 & 2141.45 &  2.21 & 0.40 & 0.43 \\
 \hline
\end{tabular}
\end{table*}
\subsection{The cluster mass function}
\label{USco:IMF_MF}

We adopt the logarithmic form of the Initial Mass 
Function as originally proposed by \citet{salpeter55}:
$\xi$($\log_{10}m$) = d$n$/d$\log_{10}$($m$) $\propto$ m$^{-\alpha}$.
We converted the luminosity into a mass using the BT-Settl 
models \citep{allard12} and the $J$-band filter for all sources. 
We assumed a distance of 145 pc \citep{vanLeeuwen09} and an age of 5 Myr 
for USco (Table \ref{tab_USco:LF_MF_5Myr}). We also considered an age 
of 10 Myr to compute the mass function (Table \ref{tab_USco:LF_MF_10Myr}),
keeping the same magnitude bins as a starting point.

We compare in Fig.\ \ref{fig_USco:LF_and_MF} the mass function derived
from our $ZYJHK-$PM sample with the previous sample of spectroscopic
members presented in \citet{lodieu11a}. We emphasise that the first and
last bins are incomplete because of saturation on the bright side and 
incompleteness on the faint side. Both mass functions were computed 
assuming an age of 5 Myr but the earlier determination made use of the
NextGen \citep{baraffe98} and DUSTY \citep{chabrier00c} whereas we
use the BT-Settl \citep{allard12} with the new dataset presented here.
Nonethless, we observe that both mass functions are similar, with
a possible excess of brown dwarfs below $\sim$0.03 M$_{\odot}$ as
originally claimed by \citet{preibisch01} and \citet{slesnick08}.
This excess of substellar objects is enhanced if we consider an
age of 10 Myr although it may be the result of the mass-luminosity
relation, which does not reproduce the M7/M8 gap proposed by
\citet{dobbie02b} occurring around 0.015--0.02 M$_{\odot}$ in USco
(left panel in Fig.\ \ref{fig_USco:LF_and_MF}).
Masses of brown dwarfs at 10 Myr are higher than at 5 Myr, making 
the number of substellar objects piling up at higher masses, causing
an enhanced bump in the mass function. We computed the mass function
using ages of 3 Myr and 1 Myr and found that the excess of brown dwarfs
seems to disappear only if the association is 1 Myr, which is lower
than any age estimate from previous {\bf{studies}}. We observe that
the high-mass part ($>$0.03 M$_{\odot}$) of the USco mass function
is well reproduced by the log-normal form of the field mass function
\citep{chabrier03,chabrier05a}, independently of the age chosen for USco
(see Section \ref{USco:IMF_age}).

We investigated the role of the bin size on the shape of the mass
function, assuming an age of 5 Myr and a distance of 145 pc. We considered
two options: on the one hand, we binned the mass function by a factor of
two i.e.\ six bins instead of 12 (open triangles in 
Fig.\ \ref{fig_USco:LF_and_MF}), and, on the other hand, we employed
four bins (open diamonds in Fig.\ \ref{fig_USco:LF_and_MF}) after 
removing the first and last bins which are incomplete
(Table \ref{tab_USco:LF_MF_5Myr}). We conclude that, 
overall, the bin size does not affect the shape of the mass function 
and the possible presence of the excess of low-mass brown dwarfs.

\subsection{Comparison with other clusters}
\label{USco:IMF_compare}

\begin{figure}
   \includegraphics[width=\linewidth]{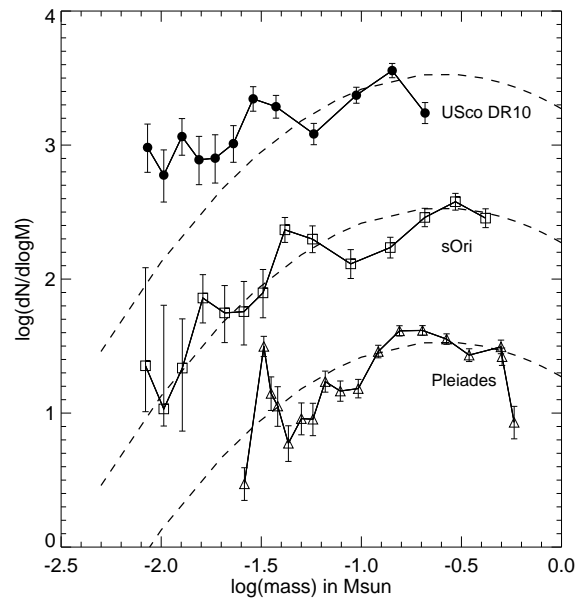}
   \caption{Mass function of USco derived from our  GCS DR10 sample 
compared to the Pleiades \citep[red open triangles;][]{lodieu12a}
and $\sigma$\,Orionis \citep[blue open squares;][]{lodieu09e}.
Overplotted as a black dashed line is the log-normal field mass function
\citep{chabrier05a}.
} 
   \label{fig_USco:MFcompare}
\end{figure}

Figure \ref{fig_USco:MFcompare} compares the USco mass function with
our studies of the Pleiades \citep[red open triangles;][]{lodieu12a} 
and $\sigma$\,Orionis \citep[blue open squares;][]{lodieu09e}. We note
that the USco mass function is derived using the latest BT-Settl models
\citep{allard12} where former studies of the Pleiades and
$\sigma$\,Orionis employed the NextGen \citep{baraffe98} and
DUSTY \citep{chabrier00c} models. The mass
function of the Pleiades comes from {\bf{a}} photometric and astrometric
selection using GCS DR9 in the same manner as this work in USco.
Both mass functions are very comparable in the interval where they
overlap, from $\sim$0.2 M$_{\odot}$ down to $\sim$0.03 M$_{\odot}$
and match the log-normal form of the field mass function 
\citep{chabrier05a}. We note that the Pleiades mass function is
comparable to the mass functions in $\alpha$\,Per \citep{lodieu12c}
and Praesepe \citep{boudreault12} using the same GCS DR9 database
in an homogeneous manner. This result is in line with the numerous
mass functions plotted in Figure 3 of \citet{bastian10}, demonstrating
the similarities between mass functions in many clusters over a broad
range of masses.

We plot in Figure \ref{fig_USco:MFcompare}
the mass function for the young (1--8 Myr) $\sigma$\,Ori cluster
derived from a pure photometric study using the fourth data release of 
the GCS \citep[blue open squares;][]{lodieu09e}. The shape of the 
$\sigma$\,Ori mass function agrees with the field mass function in the
0.2--0.01 M$_{\odot}$ mass range, as noted by independent studies of 
the cluster \citep{bejar01,caballero07d,bihain09,bejar11,penya12a}.

%
%
%

%
%
\section{Summary}
\label{USco:summary}

We have presented the outcome of a deep and wide photometric and proper 
motion survey in the USco association as part of the UKIDSS GCS DR10\@.
The main results of our analysis are:
\begin{itemize}
\item we recovered several hundred known USco members and updated 
their membership with the proper motion and photometry available 
from GCS DR10.
\item we selected photometrically and astrometrically new potential USco
member candidates and identified about 700 candidates within regions free 
of extinction and regions affected by reddening.
\item we derived the luminosity function in the USco association in the 
$J$\,=\,11.5--17.5 mag range.
\item we derived the USco mass function which matches well the log-normal 
form of the system field mass function down to 0.03 M$_{\odot}$. The
USco mass function is consistent with the Pleiades, $\alpha$\,Per, and
Praesepe mass functions in the 0.2--0.03 M$_{\odot}$ mass range.
We observe a possible excess of substellar members below 0.03 M$_{\odot}$,
as pointed out by earlier studies \citep{preibisch01,lodieu07a,slesnick08},
which may be due to the uncertainties on the mass-luminosity relation
at the M/L transition and the age of the association.
\end{itemize}

This paper provides a full catalogue of photometric and astrometric members
from 0.2 M$_{\odot}$ down to $\sim$0.01 M$_{\odot}$ in the southern part 
of the USco association. This catalogue is complemented by a complete 
census of the stellar and substellar populations in full association 
USco down to 0.02--0.015 M$_{\odot}$. This work will represent a 
reference for many years to come. We foresee further improvement when 
a second epoch will be obtained for the northern part of the association 
as part of the VISTA hemisphere survey \citep{emerson04,dalton06}. 
Our study provides a legacy sample that can be used to study the disk 
properties of low-mass stars and brown dwarfs in USco 
\citep{carpenter06,scholz07a,riaz09b,dawson12,luhman12c},
their binary properties \citep{biller11}, and their distribution in 
the association as {\bf{a}} function of spectral type once a complete
spectroscopic follow-up is available.

%
%
\section*{Acknowledgments}

NL is funded by the Ram\'on y Cajal fellowship number 08-303-01-02 and the 
national program AYA2010-19136 funded by the Spanish ministry of science and 
innovation. {\bf{I thank the anonymous referee for her/his constructive and
quick report.}} This work is based in part on data obtained as part of the UKIRT 
Infrared Deep Sky Survey (UKIDSS). The UKIDSS project is defined in 
\citet{lawrence07}. UKIDSS uses the UKIRT Wide Field Camera
\citep[WFCAM;][]{casali07}. The photometric system is described in
\citet{hewett06}, and the calibration is described in \citet{hodgkin09}.
The pipeline processing and science archive are described in \citet{irwin04}
and \citet{hambly08}, respectively.
We thank our colleagues at the UK Astronomy Technology Centre, the Joint 
Astronomy Centre in Hawaii, the Cambridge Astronomical Survey and Edinburgh 
Wide Field Astronomy Units for building and operating WFCAM and its 
associated data flow system. We are grateful to France Allard for placing her
latest BT-Settl models on a free webpage for the community.

This research has made use of the Simbad database, operated at the Centre de 
Donn\'ees Astronomiques de Strasbourg (CDS), and of NASA's Astrophysics Data 
System Bibliographic Services (ADS).

This publication makes use of data products from the Two Micron 
All Sky Survey (2MASS), which is a joint project of the University 
of Massachusetts and the Infrared Processing and Analysis 
Center/California Institute of Technology, funded by the National 
Aeronautics and Space Administration and the National Science Foundation.

%
%
\bibliographystyle{mn2e}
\bibliography{../../AA/mnemonic,../../AA/biblio_old}

%
%
\appendix

%
%
\section{USco member candidates in the $ZYJHK-$PM region free
of reddening}

\begin{table*}
  \caption{Sample of 201 USco member candidates in the GCS 
  DR10 PM region devoid of reddening, including
known members previously published in the literature. This table is 
available electronically in the online version of the journal.
}
  \label{tab_USco:candidates_GCSDR10_noExt}
  \begin{tabular}{@{\hspace{0mm}}c @{\hspace{2mm}}c @{\hspace{2mm}}c @{\hspace{2mm}}c @{\hspace{2mm}}c @{\hspace{2mm}}c @{\hspace{2mm}}c @{\hspace{2mm}}c @{\hspace{2mm}}c @{\hspace{2mm}}c @{\hspace{2mm}}c@{\hspace{0mm}}}
  \hline
R.A.\ & Dec.\  &  $Z$\,$\pm$\,err  &  $Y$\,$\pm$\,err  &  $J$\,$\pm$\,err  &  $H$\,$\pm$\,err  & $K$1\,$\pm$\,err & $K$2\,$\pm$\,err & $\mu_{\alpha}cos\delta$\,$\pm$\,err & $\mu_{\delta}$\,$\pm$\,err & $\chi^{2}$ \cr
 \hline
15:41:26.54 & $-$26:13:25.5 & 15.649$\pm$0.005 & 14.788$\pm$0.003 & 13.999$\pm$0.003 & 13.406$\pm$0.003 & 12.971$\pm$0.002 & 12.986$\pm$0.002 &    $-$7.31$\pm$2.70 &   $-$17.66$\pm$2.70 &  0.58 \cr
15:41:54.34 & $-$25:12:43.6 & 15.588$\pm$0.005 & 99.999$\pm$99.999 & 14.435$\pm$0.004 & 13.729$\pm$0.003 & 13.422$\pm$0.003 & 13.457$\pm$0.003 &    $-$8.05$\pm$1.97 &   $-$24.10$\pm$1.97 &  4.28 \cr
 \ldots{}   & \ldots{}   & \ldots{} & \ldots{} & \ldots{} & \ldots{} & \ldots{} & \ldots{}  & \ldots{}  & \ldots{} & \ldots{} \cr
16:37:05.23 & $-$26:25:44.2 & 14.266$\pm$0.003 & 13.605$\pm$0.002 & 12.963$\pm$0.002 & 12.461$\pm$0.002 & 12.082$\pm$0.001 & 12.088$\pm$0.001 &    $-$7.65$\pm$2.69 &   $-$20.81$\pm$2.69 &  0.68 \cr
16:37:54.43 & $-$26:51:52.0 & 12.968$\pm$0.001 & 12.507$\pm$0.001 & 11.923$\pm$0.001 & 11.459$\pm$0.001 & 11.096$\pm$0.001 & 11.142$\pm$0.001 &    $-$6.53$\pm$2.69 &   $-$22.52$\pm$2.69 &  0.84 \cr
 \hline
\end{tabular}
\end{table*}

\section{USco member candidates in the $ZYJHK-$PM region 
affected by reddening}

\begin{table*}
  \caption{Sample of 120 USco member candidates in the GCS
DR10 PM region affected by reddening, including
known members previously published in the literature. This table is 
available electronically in the online version of the journal.
}
  \label{tab_USco:candidates_GCSDR10_Extinction}
  \begin{tabular}{@{\hspace{0mm}}c @{\hspace{2mm}}c @{\hspace{2mm}}c @{\hspace{2mm}}c @{\hspace{2mm}}c @{\hspace{2mm}}c @{\hspace{2mm}}c @{\hspace{2mm}}c @{\hspace{2mm}}c @{\hspace{2mm}}c @{\hspace{2mm}}c@{\hspace{0mm}}}
  \hline
R.A.\ & Dec.\  &  $Z$\,$\pm$\,err  &  $Y$\,$\pm$\,err  &  $J$\,$\pm$\,err  &  $H$\,$\pm$\,err  & $K$1\,$\pm$\,err & $K$2\,$\pm$\,err & $\mu_{\alpha}cos\delta$\,$\pm$\,err & $\mu_{\delta}$\,$\pm$\,err & $\chi^{2}$ \cr
 \hline
16:16:20.73 & $-$25:17:30.1 & 13.316$\pm$0.002 & 12.800$\pm$0.001 & 12.192$\pm$0.001 & 11.528$\pm$0.001 & 11.171$\pm$0.001 & 11.231$\pm$0.001 &    $-$4.56$\pm$1.82 &   $-$16.57$\pm$1.82 &  0.53 \cr
16:16:30.68 & $-$25:12:20.3 & 14.208$\pm$0.003 & 13.549$\pm$0.002 & 12.887$\pm$0.001 & 12.320$\pm$0.002 & 11.935$\pm$0.001 & 11.936$\pm$0.001 &   $-$12.30$\pm$1.82 &   $-$19.96$\pm$1.82 &  0.48 \cr
 \ldots{}   & \ldots{}   & \ldots{} & \ldots{} & \ldots{} & \ldots{} & \ldots{} & \ldots{}  & \ldots{}  & \ldots{} & \ldots{} \cr
16:39:49.31 & $-$24:33:10.1 & 14.232$\pm$0.003 & 13.771$\pm$0.002 & 13.169$\pm$0.002 & 12.501$\pm$0.001 & 12.267$\pm$0.001 & 12.266$\pm$0.001 &   $-$12.25$\pm$1.83 &   $-$21.05$\pm$1.83 &  0.54 \cr
16:40:47.51 & $-$24:04:38.9 & 19.594$\pm$0.069 & 18.269$\pm$0.030 & 17.073$\pm$0.022 & 15.881$\pm$0.014 & 15.364$\pm$0.015 & 15.334$\pm$0.010 &    $-$7.17$\pm$2.38 &   $-$17.62$\pm$2.38 &  8.88 \cr
 \hline
\end{tabular}
\end{table*}

\section{USco member candidates in the GCS SV}

\begin{table*}
  \caption{Sample of 81 USco member candidates identified in the GCS
SV area, including previously-known members. The proper motion 
measurements come from the 2MASS/GCS cross-match. This table is
available electronically in the online version of the journal.
}
  \label{tab_USco:candidates_GCSDR10_SV}
  \begin{tabular}{@{\hspace{0mm}}c @{\hspace{2mm}}c @{\hspace{2mm}}c @{\hspace{2mm}}c @{\hspace{2mm}}c @{\hspace{2mm}}c @{\hspace{2mm}}c @{\hspace{2mm}}c @{\hspace{2mm}}c@{\hspace{0mm}}}
  \hline
R.A.\ & Dec.\  &  $Z$\,$\pm$\,err  &  $Y$\,$\pm$\,err  &  $J$\,$\pm$\,err  &  $H$\,$\pm$\,err  & $K$1\,$\pm$\,err & $\mu_{\alpha}cos\delta$ & $\mu_{\delta}$ \cr
 \hline
16:06:03.75 & $-$22:19:30.0 & 18.169$\pm$0.036 & 16.825$\pm$0.014 & 15.853$\pm$0.009 & 15.096$\pm$0.009 & 14.438$\pm$0.009 &    13.67 &   $-$26.51 \cr
16:06:06.29 & $-$23:35:13.3 & 18.430$\pm$0.041 & 17.150$\pm$0.018 & 16.204$\pm$0.012 & 15.540$\pm$0.012 & 14.973$\pm$0.012 &    $-$6.08 &     5.64 \cr
 \ldots{}   & \ldots{}   & \ldots{} & \ldots{} & \ldots{} & \ldots{} & \ldots{} & \ldots{}  & \ldots{}  \cr
16:17:01.47 & $-$23:29:06.0 & 13.656$\pm$0.002 & 13.025$\pm$0.002 & 12.453$\pm$0.001 & 11.887$\pm$0.001 & 11.537$\pm$0.001 &   $-$17.93 &   $-$19.58 \cr
16:17:06.06 & $-$22:25:41.6 & 13.254$\pm$0.002 & 12.663$\pm$0.002 & 12.120$\pm$0.001 & 11.516$\pm$0.001 & 11.202$\pm$0.001 &   $-$15.03 &   $-$26.36 \cr
 \hline
\end{tabular}
\end{table*}

\section{USco member candidates in the $HK$ sample}

\begin{table*}
  \caption{Sample of 286 USco member candidates identified in the USco
region imaged in $H$ and $K$ only, including previously-known members. 
This table is available electronically in the online version of the journal.
}
  \label{tab_USco:candidates_GCSDR10_HKonly}
  \begin{tabular}{@{\hspace{0mm}}c @{\hspace{2mm}}c @{\hspace{2mm}}c @{\hspace{2mm}}c @{\hspace{2mm}}c @{\hspace{2mm}}c @{\hspace{2mm}}c@{\hspace{0mm}}}
  \hline
R.A.\ & Dec.\  &  $H$\,$\pm$\,err  & $K$\,$\pm$\,err &  $\mu_{\alpha}cos\delta$ & $\mu_{\delta}$ \cr
 \hline
15:40:10.22 & $-$24:31:18.4 & 14.538$\pm$0.004 & 13.818$\pm$0.004 &     0.31 &    $-$9.96 \cr
15:41:16.03 & $-$25:30:56.5 & 14.932$\pm$0.008 & 14.305$\pm$0.008 &    $-$9.04 &    $-$7.19 \cr
 \ldots{} & \ldots{} & \ldots{} & \ldots{} & \ldots{}  &  \ldots{} \cr
16:40:19.60 & $-$22:18:12.0 & 13.361$\pm$0.002 & 12.953$\pm$0.002 &     0.36 &    $-$5.89 \cr
16:40:24.95 & $-$22:18:52.8 & 12.871$\pm$0.002 & 12.450$\pm$0.002 &     5.38 &   $-$13.56 \cr
 \hline
\end{tabular}
\end{table*}

\section{USco member candidates With WISE photometry}

\begin{table*}
  \caption{Sample of 660 USco member candidates identified in GCS DR10
with WISE photometry, ordered by increasing right ascension. 
The coordinates are from the UKIDSS GCS DR10 database whereas the 
photometry is from the WISE all-sky release. Objects detected in two, 
three, and four WISE bands are marked as 1100, 1110, and 1111
(column 7; wise\_detect), respectively. This table is available 
electronically in the online version of the journal.
}
  \label{tab_USco:candidates_GCSDR10_WISE}
  \begin{tabular}{@{\hspace{0mm}}c @{\hspace{2mm}}c @{\hspace{2mm}}c @{\hspace{2mm}}c @{\hspace{2mm}}c @{\hspace{2mm}}c @{\hspace{2mm}}c @{\hspace{2mm}}c@{\hspace{0mm}}}
  \hline
R.A.\ & Dec.\  &  $w1$\,$\pm$\,err  & $w2$\,$\pm$\,err &  $w3$\,$\pm$\,err & $w4$\,$\pm$\,err & wise\_detect \cr
 \hline
15:40:10.22 & $-$24:31:18.4 & 13.733$\pm$0.031 & 13.829$\pm$0.048 & 12.462$\pm$0.000 &  8.811$\pm$0.000 & 1100 \cr
15:41:16.03 & $-$25:30:56.5 & 14.226$\pm$0.034 & 14.330$\pm$0.079 & 12.185$\pm$0.000 &  8.764$\pm$0.000 & 1100 \cr
 \ldots{} & \ldots{} & \ldots{} & \ldots{} & \ldots{}  &  \ldots{} & \ldots{} \cr
16:40:24.95 & $-$22:18:52.8 & 12.113$\pm$0.024 & 12.012$\pm$0.028 & 11.473$\pm$0.278 &  7.397$\pm$0.117 & 1100 \cr
16:40:47.51 & $-$24:04:38.9 & 15.028$\pm$0.054 & 14.773$\pm$0.093 & 11.426$\pm$0.000 &  8.249$\pm$0.000 & 1100 \cr
 \hline
\end{tabular}
\end{table*}

\end{document}